\documentclass[letterpaper,twocolumn,10pt]{article}
\usepackage{usenix-2020-09}

\usepackage{tikz}
\usepackage{amsmath}
\usepackage{amssymb}
\usepackage{amsthm}
\usepackage{graphicx}
\setkeys{Gin}{width=\textwidth, height=\textheight, keepaspectratio}
\graphicspath{{images/}}
\usepackage[nobiblatex]{xurl} %
\usepackage{adjustbox}
\usepackage{multirow}
\usepackage{booktabs}
\usepackage[n,operators,advantage,sets,adversary,landau,probability,notions,logic,ff,mm,primitives,events,complexity,asymptotics,keys]{cryptocode}
\usepackage{xspace}

\usepackage[all]{xy}
\usepackage{algorithm}
\usepackage{flushend}
\usepackage{paralist}
\usepackage{algpseudocode}
\makeatletter
\algnewcommand{\LineComment}[1]{\Statex \hskip\ALG@thistlm \(\triangleright\) #1}
\makeatother

\usepackage[hang,flushmargin,stable]{footmisc}

\renewcommand{\footnoterule}{%
  \kern -3pt
  \hrule width 1in
  \kern 2pt
}

\theoremstyle{definition}
\newtheorem{definition}{Definition}
\newtheorem{theorem}{Theorem}

\usepackage{url}
\makeatletter
\def\url@leostyle{%
  \@ifundefined{selectfont}{\def\UrlFont{}}%
  {\def\UrlFont{}}%
}
\makeatother
 \hypersetup{
 	colorlinks,
   linkcolor={green!40!black},
   citecolor={red!50!black},
   urlcolor={blue!60!black},
 }

\let\OLDthebibliography\thebibliography
\renewcommand\thebibliography[1]{
  \OLDthebibliography{#1}
  \setlength{\parskip}{2pt}
  \setlength{\itemsep}{0pt plus 0.3ex}
}

\newcommand{\reduce}{\vspace{-0.15cm}}

\usepackage{subcaption}

\newcommand{\descr}[1]{\smallskip\noindent\textbf{#1}}
\newcommand{\mywidth}{0.99}
\newcommand{\empeps}{\varepsilon_{emp}}
\newcommand{\approxdp}{($\varepsilon, \delta$)-DP\xspace}
\newcommand{\gdp}{$\mu$-GDP\xspace}

\newif\iflong
\longtrue
\longfalse
\iflong
	\newcommand{\longVer}[1]{#1}
	\newcommand{\shortVer}[1]{}
\else
	\newcommand{\longVer}[1]{}
	\newcommand{\shortVer}[1]{#1}	
\fi
	
\usepackage{titlesec}
\titlespacing*{\section}{0pt}{*2.5}{*2.5}
\titlespacing{\subsection}{0pt}{*2}{*2}
\titlespacing{\subsubsection}{0pt}{*2}{*2}
\usepackage{indentfirst}

\usepackage{etoolbox}
\makeatletter
\pretocmd\@maketitle
  {\def\@makefnmark{\hbox{\@textsuperscript{\normalfont\@thefnmark}}}}
  {}{\FAIL}
\makeatother

\begin{document}

\date{}

\title{\Large ``What do you want from theory alone\footnotemark?'' Experimenting with Tight Auditing of Differentially Private Synthetic Data Generation\footnotemark}

\author{
    \rm Meenatchi Sundaram Muthu Selva Annamalai$^1$, Georgi Ganev$^{1,2}$, Emiliano De Cristofaro$^3$\\[1ex]
    $^1$University College London$\;\;$ $^2$Hazy$\;\;$ $^3$University of California, Riverside
}

\maketitle

\renewcommand*{\thefootnote}{\fnsymbol{footnote}}
\footnotetext{$^*$See \url{https://www.imdb.com/title/tt15398776/quotes/?item=qt6969434}.}
\footnotetext{$^\dag$To appear at USENIX Security 2024.}

\renewcommand*{\thefootnote}{\arabic{footnote}}
\setcounter{footnote}{0}

\begin{abstract}
Differentially private synthetic data generation (DP-SDG) algorithms are used to release datasets that are structurally and statistically similar to sensitive data while providing formal bounds on the information they leak.
However, bugs in algorithms and implementations may cause the actual information leakage to be higher.
This prompts the need to verify whether the theoretical guarantees of state-of-the-art DP-SDG implementations also hold in practice.
We do so via a rigorous {\em auditing} process: we compute the information leakage via an adversary playing a distinguishing game and running membership inference attacks (MIAs). 
If the leakage observed empirically is higher than the theoretical bounds, we identify a DP violation; if it is non-negligibly lower, the audit is loose. 

We audit six DP-SDG implementations using different datasets and threat models and find that black-box MIAs commonly used against DP-SDGs are severely limited in power, yielding remarkably loose empirical privacy estimates.
We then consider MIAs in stronger threat models, i.e., passive and active white-box, using both existing and newly proposed attacks.
Overall, we find that, currently, we do not only need white-box MIAs but also worst-case datasets to {\em tightly} estimate the privacy leakage from DP-SDGs.
Finally, we show that our automated auditing procedure finds both known DP violations (in 4 out of the 6 implementations) as well as a new one in the DPWGAN implementation that was successfully submitted to the NIST DP Synthetic Data Challenge.

The source code needed to reproduce our experiments is available from \url{https://github.com/spalabucr/synth-audit}.
\end{abstract}

\section{Introduction}

\begin{table}[t]
  \centering
  \small
  \begin{tabular}{lll}
  \toprule
  {\bf Method} & {\bf Threat Model} &  {\bf Violation}\\
  {\bf (Implementation)} &  & \\
  \midrule
	PrivBayes (DS) & Black-box & Metadata \\
	PrivBayes (DS) & White-box & Pre-processing \\
  PrivBayes (Hazy) & Black-box & Metadata \\
  MST (Smartnoise) & Black-box & Metadata \\
  DPWGAN (NIST) & Active White-box & Early stopping \\
	DPWGAN (Synthcity) & Black-box & Metadata\\
	DPWGAN (Synthcity) & Black-box & PRNG Reuse\\
  \bottomrule
  \end{tabular}
  \caption{Overview of identified privacy violations.}\label{tab:intro_summary}
\end{table}

In modern data-driven ecosystems, organizations are often compelled or willing to share data within and with each other~\cite{eu2022datastrategy}.
However, even if it is ``sanitized,'' ``anonymized,'' or aggregated, sharing data can still lead to severely violating the privacy of the data subjects~\cite{uscensusreconstruction}.
In this context, Synthetic Data Generation (SDG) algorithms have been proposed as a potential mitigation; by learning the underlying distribution of the sensitive data and then sampling ``fresh'' synthetic data points from it, SDGs enable entities to generate and release artificial data that, ostensibly, only statistically resembles the real data.
However, without formal privacy protections, SDGs can easily leak sensitive user data~\cite{stadler2022synthetic,hayes2017logan,houssiau2022tapas,yale2019assessing,annamalai2023linear}.

The standard, rigorous way to define algorithms with formally bounded information leakage is through Differential Privacy (DP)~\cite{dwork2006calibrating}.
Researchers have proposed a number of SDGs that satisfy DP, aka DP-SDGs~\cite{mckenna2021winning,zhang2017privbayes,xie2018differentially,jordon2018pate,cai2021data,aydore2021differentially,liu2021iterative}.
Particularly in the tabular data domain~\cite{mckenna2021winning,zhang2017privbayes,li2021dpsyn}, DP-SDGs have started to see real-world adoption; e.g., in 2021, Microsoft and the UN International Organization for Migration released a (DP) synthetic dataset that describes victim-perpetrator relations in the context of human trafficking.

However, bugs that lead to DP violations have been found in several popular DP tools~\cite{nasr2023tight,bichsel2021dp,lokna2023group}, including in tabular DP-SDGs~\cite{lokna2023group,stadler2022synthetic}.
This motivates the need to {\em audit} state-of-the-art DP-SDG implementations, i.e., 
designing and executing experiments to derive {\em empirical} privacy leakage estimates.
These are then compared against the {\em theoretical} (provable) DP guarantees to verify the correctness of implementations~\cite{lokna2023group} and/or detect DP violations~\cite{jagielski2020auditing,jayaraman2019evaluating,nasr2023tight}.

The auditing process often relies on membership inference attacks (MIAs), where an adversary attempts to learn whether or not a given record was used as input to the algorithm, following a distinguishing game meant to mirror the DP definition~\cite{stadler2022synthetic}. %
The threat models in which MIAs can be mounted include what, in Section~\ref{sec:threat}, we define as black- and white-box settings; in the former, the adversary only has access to the synthetic data, while, in the latter, she also sees the trained generative model and its internal parameters.

Prior work studying privacy in DP-SDGs has mostly focused on {\em black-box} attacks and used randomly sampled (\textit{average-case}) training datasets~\cite{stadler2022synthetic}.
Alas, these are likely to yield {\em loose} empirical bounds~\cite{houssiau2022tapas}, i.e., the empirical estimates are not close to the theoretical DP guarantees.
Conversely, prior work auditing DP algorithms in discriminative settings
has obtained tight estimates by considering {\em active} white-box~\cite{nasr2018comprehensive} attacks, where the adversary can manipulate the training process by inserting arbitrary canary gradients~\cite{nasr2021adversary,nasr2023tight}.

In this work, we present a comprehensive audit of DP-SDGs.
This prompts two main research questions: 
{\em 1)} How {\em tightly} can we empirically estimate privacy leakage in DP-SDGs?
{\em 2)} How do different threat models and %
datasets affect tightness?
To this end, we design an experimental framework including MIAs against different DP-SDG implementations, training datasets, and threat models. %
Using our framework, we audit three state-of-the-art tabular DP-SDG algorithms (PrivBayes~\cite{zhang2017privbayes}, MST~\cite{mckenna2021winning}, and  DPWGAN~\cite{dpwgannist}), considering two independent implementations for each algorithm. %

\descr{Main Findings.} 
Our analysis shows that: 
\begin{itemize}
  \item Common black-box MIAs like the distance to closest record (DCR) heuristic %
  are ineffective at exploiting privacy leakage from DP-SDGs. %
  \item White-box and active white-box attacks provide much tighter empirical privacy estimation, especially with specially crafted \textit{worst-case} datasets. For instance, for MST at theoretical $\varepsilon = 4.0$, white-box auditing produces empirical privacy estimation of $\empeps = 3.10$ compared to black-box's meaningless estimates ($\empeps = 0.00$).
  \item The tightest possible settings may be implementation-dependent, i.e., we need different \textit{worst-case} datasets and threat models to achieve tight empirical privacy estimates for different DP-SDG implementations. 
  E.g., {\em passive} white-box audits of PrivBayes and MST are tight, while DPWGAN requires {\em active} white-box attacks.
  \item As summarized in Table~\ref{tab:intro_summary}, we find DP violations in four out of the six implementations we study, due to learning metadata directly from the input. 
  We also identify a new DP violation in the DPWGAN implementation successfully submitted to the NIST DP Synthetic Data Challenge~\cite{dpsynth2018}.
\end{itemize}

\descr{Contributions.} In summary, our main contributions include:
\begin{enumerate}
  \item We perform the first large-scale audit of DP-SDG algorithms and their implementations.
  \item We craft implementation-specific worst-case datasets for DP-SDGs, which enables us to achieve {\em tight} audits.
  \item We present the first white-box MIAs against PrivBayes and MST.
\end{enumerate} 

\longVer{Our work offers a re-usable DP-SDG auditing framework and enables reasoning about privacy guarantees vis-\`a-vis different settings and threat models.
We are confident it will encourage further research into automated auditing tools to verify DP-SDG implementations easily and at scale.
}

\section{Preliminaries}
We now introduce the concepts of differential privacy, auditing, membership inference, and synthetic data generation.

\subsection{Differential Privacy (DP)}
\label{sec:prelims_dp}

\begin{definition}[Differential Privacy (DP)~\cite{dwork2006calibrating}]
  \label{def:dp}
  A randomized mechanism $\mathcal{M} : \mathcal{D} \rightarrow \mathcal{R}$ is $(\varepsilon, \delta)$-differentially private if for any two neighboring datasets $D, D' \in \mathcal{D}$ and $S \subseteq \mathcal{R}$, it holds: %
  \begin{equation*}
    \Pr[\mathcal{M}(D) \in S]  \leq e^\varepsilon \Pr[\mathcal{M}(D') \in S] + \delta %
  \end{equation*}
\end{definition}

\noindent Definition~\ref{def:dp} describes the so-called \textit{approximate} DP variant, which is a relaxation of the original (``{\em pure}'') DP definition whereby $\delta = 0$.
We also consider two variants of DP depending on the definition of {\em neighboring datasets}: 1) {\em add/remove}, aka unbounded, and 2) {\em edit}, aka bounded, DP.
The former corresponds to inserting/deleting a single record from the dataset ($|D| = |D'| \pm 1$); the latter entails replacing a single record with another ($|D| = |D'|$).

An important property of DP is given by the \textit{post-processing} theorem, which lets us use the output of DP mechanisms freely without worrying about further privacy leakage.\reduce %

\begin{theorem}[Post-Processing]
  Let $\mathcal{M}: \mathcal{D} \rightarrow \mathcal{R}$ be an \approxdp mechanism and $f: \; \mathcal{R} \rightarrow \mathcal{R}'$. Then $f \circ \mathcal{M}: \mathcal{D} \rightarrow \mathcal{R}'$ also satisfies $(\epsilon, \delta)$-DP.
\end{theorem}

\subsection{Auditing DP}
\label{sec:prelims_audit_dp}
Implicit to the DP definition is a theoretical limit on any adversary's ability to distinguish between outputs of an \approxdp mechanism $\mathcal{M}$ (i.e., $\mathcal{M}(D)$ and $\mathcal{M}(D')$).
When observed in practice, this limit can be used to estimate the empirical guarantees provided by a DP mechanism.
The process of \textit{auditing} DP entails verifying the \textit{theoretical} guarantees provided by $\mathcal{M}$ by running an experiment where an adversary attempts to distinguish between $\mathcal{M}(D)$ and $\mathcal{M}(D')$ and estimating the \textit{empirical} guarantees ($\empeps, \delta$) from the adversary's success.

Informally, when auditing a DP mechanism $\mathcal{M}$, $\mathcal{M}$ is repeatedly run on a pair of fixed neighboring datasets $D$ and $D'$ to generate two sets of observations $\mathcal{O} = \{o_1, o_2, ...\}$ and $\mathcal{O}' = \{o'_1, o'_2, ...\}$, respectively.
Next, an adversary attempts to distinguish between the two sets of outputs, which results in a false positive rate $\alpha$ and a false negative rate $\beta$.
(We provide a formal definition in Section~\ref{sec:dp_game}).
Then, upper bounds $\overline{\alpha}$ and $\overline{\beta}$ can be calculated using Clopper-Pearson confidence intervals, as done in previous work~\cite{nasr2021adversary,nasr2023tight}.
Finally, the upper bounds on $\alpha$ and $\beta$ are converted back into an empirical lower bound $\empeps$ using two known methods, i.e., auditing using either the \approxdp or the \gdp definition, described below.

\descr{Maximum auditable $\varepsilon$.} The empirical lower bound, $\empeps$, has a confidence level that follows the upper bounds' confidence level.
At the same time, this imposes an inherent \textit{limit} on the maximum $\empeps$ that can be derived in this way, which we refer to as the \textit{maximum auditable $\varepsilon$}.
Intuitively, even if the adversary can perfectly distinguish between $O$ and $O'$ (i.e., $\alpha = \beta = 0$), $\overline{\alpha}$ and $\overline{\beta}$ are lower bounded by the confidence interval thus resulting in an upper bound on $\empeps$.

\descr{Auditing using \approxdp}
In general, any mechanism that satisfies \approxdp bounds the possible false positive rates ($\alpha$) and false negative rates ($\beta$) attainable by an adversary to the following privacy region~\cite{kairouz2015composition}: %
\begin{align*}
  \mathcal{R}(\varepsilon, \delta) = \{ (\alpha, \beta) | & \alpha + e^\varepsilon \beta \geq 1 - \delta \land e^\varepsilon \alpha + \beta \geq 1 - \delta \land \\
  & \alpha + e^\varepsilon \beta \leq e^\varepsilon + \delta \land e^\varepsilon \alpha + \beta \leq e^\varepsilon + \delta \} %
\end{align*}
An empirical lower bound $\empeps$ can be calculated according to the privacy region using the following equation: %
\begin{align}\label{eq1}
  \empeps = \max \left\{\ln \left( \frac{1 - \overline{\alpha} - \delta}{\overline{\beta}} \right), \ln \left( \frac{1 - \overline{\beta} - \delta}{\overline{\alpha}} \right), 0 \right\} %
\end{align}
When auditing pure DP, we can use Eq.~\ref{eq1} with $\delta = 0$.

\descr{Auditing using \gdp.}
While the \approxdp auditing method applies in \textit{general} to all approximate DP mechanisms, Nasr et al.~\cite{nasr2023tight} note that mechanisms can have privacy regions that are \textit{specific} to the mechanism as well.
For example, mechanisms that satisfy $\mu$-Gaussian Differential Privacy (GDP) also satisfy approximate DP but define a much smaller subset of $\mathcal{R}(\varepsilon, \delta)$ as its privacy region.
Therefore, we can audit \gdp mechanisms by first converting the bounds on $\alpha$ and $\beta$ into a lower bound on $\mu$ using the following equation: %
\begin{equation}
  \mu_{emp} = \Phi^{-1}(1 - \overline{\alpha}) - \Phi^{-1} (\overline{\beta}) %
\end{equation}
We can then convert the lower bound $\mu_{emp}$ to a lower bound $\empeps$ using the following theorem for a fixed $\delta$: \reduce
\begin{theorem}[\gdp to \approxdp conversion~\cite{dong2019gaussian}]
  A mechanism is \gdp iff it is $(\varepsilon, \delta(\varepsilon))$-DP for all $\varepsilon \geq 0$, where: %
  \begin{equation}
    \delta(\varepsilon) = \Phi \left( -\frac{\varepsilon}{\mu} + \frac{\mu}{2} \right) - e^\varepsilon \Phi \left( -\frac{\varepsilon}{\mu} - \frac{\mu}{2} \right) %
  \end{equation}
\end{theorem}

\subsection{Membership Inference Attacks (MIAs)}
In a membership inference attack (MIA), the adversary aims to determine if a target record, $x_T$, was used in input to a function -- e.g., aggregation~\cite{pyrgelis2018knock}, training a model~\cite{shokri2017membership}, etc.
In recent years, a number of MIAs against machine learning models have been presented that consider various threat models and settings~\cite{shokri2017membership,choquette2021label,hayes2017logan,ye2022enhanced}.

In the DP auditing setting, we define an MIA as a function that takes in input the two neighboring datasets ($D$, $D'$), the target record ($x_T$), the mechanism being audited ($\mathcal{M}$), and a single output $y$ of the mechanism run on $D$ or $D'$.
The MIA function outputs a (possibly unbounded) \textit{score}, $s$, that represents the confidence the attack assigns to the event that $D$ was the input to the mechanism based on the output (i.e., $y \sim \mathcal{M}(D)$).
In short, we define the MIA function as: %
\begin{equation}
s \leftarrow \text{MIA}(x_T, y; \mathcal{M}, D, D'). %
\end{equation}

\subsection{Synthetic Data Generation (SDG)}
Synthetic data generation (SDG) algorithms take in input an \textit{original} dataset $D$ and output a \textit{synthetic} dataset $S$.
Typically, a generative model $\mathcal{G}$ is first fit on $D$ using a (possibly randomized) fitting function, i.e. $\mathcal{G} \sim \text{GM}(D)$.
A synthetic dataset with $m$ records is then sampled from this model, i.e., $S \sim \mathcal{G}^m$.

In differentially private synthetic data generation algorithms (DP-SDGs), the fitting function $\text{GM}$ itself typically satisfies DP.
That is, the probability that the adversary can infer if a given generative model $\mathcal{G}$ was fit on $D$ or $D'$, i.e., $\mathcal{G} \sim \text{GM}(D)$ or $\mathcal{G} \sim \text{GM}(D')$, is bounded by the $\varepsilon$ parameter.
The guarantees of the overall SDG algorithm then follow from the post-processing theorem, as the synthetic dataset is simply sampled from the fitted generative model.
However, DP-SDGs might pre-process the dataset without proper DP accounting, which in practice can result in DP violations~\cite{stadler2022synthetic}.

\section{Auditing DP-SDG Algorithms}

\subsection{Overview}
We now set out to audit a differentially private synthetic data generation (DP-SDG) algorithm $\mathcal{M}_\text{SDG}$ using an adversary and a distinguishing game;
given neighboring datasets $D$ and $D'$, the adversary distinguishes between outputs from $\mathcal{M}_\text{SDG}(D)$ and $\mathcal{M}_\text{SDG}(D')$.
More precisely, she distinguishes using a membership inference attack (MIA).
The attack's success rate -- i.e., the number of false positives and false negatives -- is then used to compute a lower-bound empirical estimate, $\empeps$, of the privacy leakage.

We consider the auditing procedure to be ``tight'' if the empirical estimate $\empeps$ is close to the theoretical guarantee $\varepsilon$.
Thus, the auditing procedure can be used to %
identify privacy violations~\cite{lokna2023group,nasr2023tight,debenedetti2023privacy}, if $\empeps \gg \varepsilon$.
It can also be used to determine if the theoretical guarantees are loose or if there is significant room for the membership inference attacks to be improved~\cite{nasr2021adversary}, if $\empeps \ll \varepsilon$.

We instantiate a range of MIAs with varying adversarial capabilities and study the impact of auditing decisions on tightness. %
In the rest of this section, we define the threat models considered in this work, formalize the DP distinguishing game, and define and introduce the methodology used to select \textit{worst-case} target records and neighboring datasets. %

\subsection{Threat Models}\label{sec:threat}
Our analysis considers progressively stronger threat models (introduced below) to study the power of the adversary needed to achieve tight empirical estimates.
While there have been many different definitions of threat models introduced in prior work, we follow the definitions provided by Houssiau et al.~\cite{houssiau2022tapas} for the black- and white-box threat models as they are meant specifically for SDGs.
We will assume, in \textit{all} threat models, that the adversary can choose a worst-case target record and has knowledge of the neighboring datasets $D$ and $D'$, as is standard for auditing DP mechanisms~\cite{nasr2023tight,nasr2021adversary,jagielski2020auditing}.

\descr{Black-Box.} In the black-box setting, the adversary has access to the synthetic dataset, $S$, as well as to the specifications of the SDG algorithm, but, crucially, not to the trained generative model $\mathcal{G}$ the synthetic dataset is sampled from.

Although this is the most practical (i.e., the weakest) threat model we consider, it is not the adversarial capability assumed by the DP guarantees provided by DP-SDGs.
This is because the generative model fitting function, $\text{GM}(\cdot)$, typically satisfies DP.
The DP guarantees then transfer to the entire SDG algorithm as per the post-processing theorem.
Thus, they should, in theory, hold even if the adversary has access to $\mathcal{G}$, and not just $S$.
Nevertheless, many synthetic data libraries use simple black-box attacks to evaluate the privacy of differentially private synthetic data~\cite{houssiau2022tapas,qian2023synthcity}.
Hence, we include this setting to compare the effectiveness of different black-box attacks in the DP setting and evaluate the impact of reduced adversarial power on empirical privacy estimates.

\descr{Passive White-Box.} Here, we assume the adversary has access to the trained generative model $\mathcal{G}$ and its internal parameters, in addition to the synthetic dataset $S$.
Examples of MIAs in this setting include~\cite{hayes2017logan,hilprecht2019monte}.

\descr{Active White-Box.} Finally, we consider an adversary with not only access to the trained generative model but can also \textit{actively} manipulate training.
This setting was introduced by Nasr et. al.~\cite{nasr2023tight} in the context of auditing differentially private stochastic gradient descent (in discriminative models).
In other words, the active white-box adversary can insert arbitrary (``canary'') gradients into the training process of a machine learning model.

Note that while DP-SDGs like DPWGAN~\cite{xie2018differentially} and PATE-GAN~\cite{jordon2018pate} involve machine learning, they typically use different optimization algorithms rather than stochastic gradient descent (e.g., RMSProp).
Therefore, we include this threat model to audit the DP guarantees of DP-SDGs that use machine learning models (e.g., GANs, VAEs, diffusion models).

\descr{Worst-Case Dataset.} We also consider a setting where the adversary can choose worst-case neighboring datasets $D$ and $D'$.
Recall that DP guarantees hold not only for average-case neighboring datasets but also for worst-case ones~\cite{dwork2006calibrating}.
While Nasr et al.~\cite{nasr2023tight} show that, for discriminative models, tight empirical estimates can be achieved even for average-case datasets, MIAs are generally harder for generative models~\cite{hayes2017logan}, thus motivating us to consider both average and worst-case settings in our evaluation.

\subsection{DP Distinguishing Game}
\label{sec:dp_game}
\begin{figure}[t]
    \centering
    \fbox{\footnotesize
    \centering
    \begin{minipage}{0.9\columnwidth}
      \hspace{0.1cm}{\bf Game Parameters}: $(\mathcal{D}, n, m, \text{GM}, \tau)$\\[1ex]
      \begin{tabular}{lcl}
        \hspace*{-0.1cm}\textbf{\underline{Adversary}} & & \textbf{\underline{Challenger}}\\[0.5ex]
        \hspace*{-0.1cm}$[1]~\text{\bf Pick~}x_T \in \mathcal{D} $ \hspace{-0.15cm}  & $\xymatrix@1@=50pt{\ar[r]^*{x_T}&}$\hspace*{-0.25cm} &\\
        & & $D^- \sim \mathcal{D}^{\,n - 1} \text{~s.t.~} x_T \notin D^-$\\[1ex]
        \hspace*{-0.1cm}$[2]~\textbf{Pick~}D^* \in \mathcal{D}^{\,n - 1} $ \hspace{-0.5cm} & $\xymatrix@1@=50pt{\ar[r]^*{D^*}&}$\hspace*{-0.25cm} & $D^- := D^*~[2]$\\
        & & \\
        & & $D_0 := D^-~[1]$\\
        & & $D_1 := D^- \cup \{x_T\} $\\[-1ex]
        & $\xymatrix@1@=50pt{& \ar[l]_*{D_0,\, D_1}}$\hspace*{-0.25cm}&\\[1ex]
        & & $b \sim \bin $\\[1ex]
        $[3]$& \xymatrix@1@=50pt{& \ar@{<->}[l]_*{...}} & $\mathcal{G}_b \sim \text{GM}(D_b)$\\
        $[3,4]$& $\xymatrix@1@=50pt{& \ar[l]_*{\mathcal{G}_b}}$ & \hspace*{-0.25cm}\\[1ex]
        & & $S \sim \mathcal{G}^m_b$\\[-1ex]
        & $\xymatrix@1@=50pt{& \ar[l]_*{S}}$\hspace*{-0.25cm}&\\[2ex]
      \end{tabular}
      \hspace*{0.1cm}$s \leftarrow \text{MIA}(x_T, S, \mathcal{G}_b[3,4]; \text{GM}, D, D')$\\
      \hspace*{0.1cm}$\widehat{b} := \begin{cases} \text{0} &s < \tau \\ \text{1} &s \geq \tau \\ \end{cases} $\\ %
      \hspace*{0.1cm}$\text{\bf Output:~} \hat{b} $ \\[2ex]
      \footnotesize\hspace*{0.1cm}$[1]$~Will be modified to accommodate edit DP (see below).\\
      \footnotesize\hspace*{0.1cm}$[2]$~Only executed in the \textit{worst-case dataset} setting.\\
      \footnotesize\hspace*{0.1cm}$[3]$~Only executed in the \textbf{\it active white-box} setting.\\
            \footnotesize\hspace*{0.1cm}$[4]$~Only executed in the \textbf{\it white-box} setting.

    \end{minipage}
  }%
  \caption{Distinguishability Game between Adversary and Challenger for add/remove DP, given a raw dataset ($\mathcal{D}$), the number of records in the original dataset ($n$), the number of records in the synthetic dataset ($m$), the generative model fitting function ($\text{GM}$), and a decision threshold $\tau$.}
  \label{fig:dg}
\end{figure}
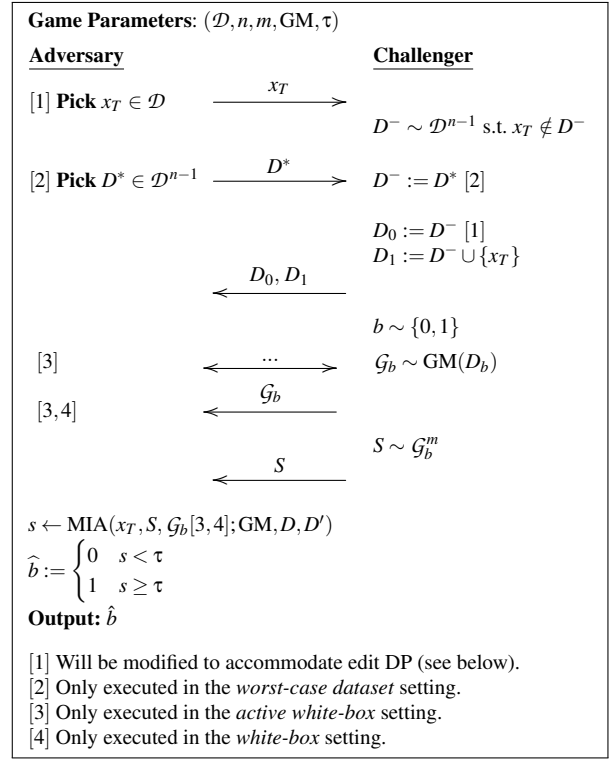

In Figure~\ref{fig:dg}, we formalize the distinguishing game, played between an Adversary and a Challenger, used to audit the (add/remove) DP guarantees of a given DP-SDG algorithm.

While the game is for the add/remove DP definition, some SDGs -- e.g., PrivBayes~\cite{zhang2017privbayes} -- satisfy the edit DP definition instead; to this end, we modify the game to audit SDG algorithms that satisfy edit DP as follows.
The Adversary chooses a worst-case pair of target records $x_T$ and $y$ instead of only $x_T$.
Next, the Challenger sets $D_0 := D^- \cup \{y\}$ rather than $D_0 := D^-$.
Doing so ensures that the Adversary distinguishes between $D^- \cup \{x_T\}$ and $D^- \cup \{y\}$.

\subsection{Worst-Case Target Record}

\begin{figure*}[t]
  \centering
  \fbox{
  \includegraphics[width=0.95\linewidth]{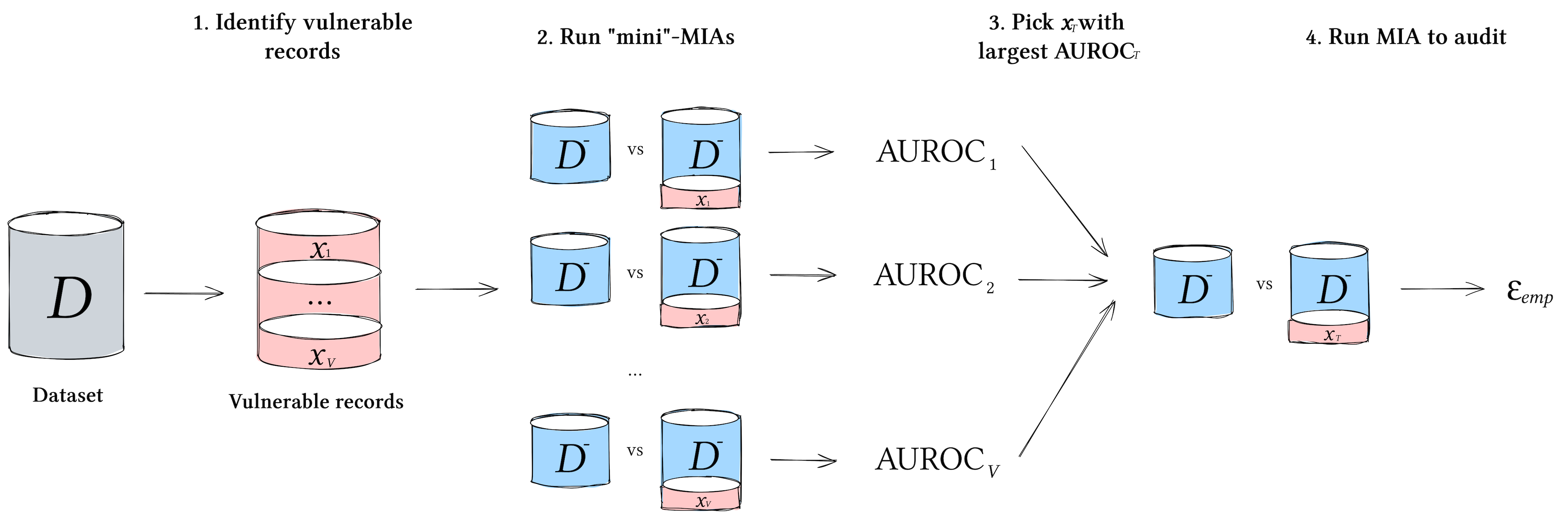}}
  \caption{Choosing the worst-case target record to audit.}
  \label{fig:choose_vuln}
\end{figure*}

As mentioned, DP provides guarantees not only for the \textit{average-case} but also \textit{worst-case}  target record.
Thus, all adversaries have the ability to choose the target record.
However, na\"ively evaluating the privacy guarantees for the worst-case record would require every possible record in the domain to be audited; in practice, this is infeasible given the number of records in common high-dimensional datasets.

Rather, we let the Adversary use the vulnerable record identification procedure by Meeus et al.~\cite{meeus2023achilles} to first select a subset of $V$ most vulnerable target records from the original dataset.
She then runs ``mini'' membership inference attacks on these $V$ records and evaluates the area under the receiver operating characteristic curve (AUC) for each record.
We call these ``mini''-MIAs as they are only run over a small number of repetitions for each record (thus, they cannot be used to audit the DP guarantees with high confidence) and use them to let the Adversary choose the record with the highest AUC as the worst-case target record during the audit.

We find that $V=100$ vulnerable records and 64 repetitions of each ``mini''-MIA are enough to identify the worst-case target record.
This process is outlined in Figure~\ref{fig:choose_vuln}.

\subsection{Worst-Case Neighboring Datasets}
\label{sec:choosing_worst_case_dataset}
Similar to the choice of target record, DP provides guarantees not only against \textit{average-case} (aka \textit{natural}) neighboring datasets but also against the \textit{worst-case} pair.
While attacks are typically evaluated against the former~\cite{shokri2017membership,hayes2017logan,houssiau2022tapas}, DP violations may not necessarily occur in these settings.
Furthermore, leakage from target records may not be maximized, leading to loose empirical estimates that are far from the theoretical guarantees.
Therefore, when auditing DP, it makes sense also to consider the worst-case neighboring datasets~\cite{nasr2021adversary,lokna2023group}.

However, worst-case neighboring datasets are likely algorithm-dependent, as leakage from different algorithms might be maximized in different settings.
Designing worst-case datasets may also not be trivial in practice as there may be edge-case inputs the program may fail to execute.
While this behavior technically constitutes a DP violation in itself, the audit, similar to error handling in compilers, should aim to identify as many DP violations as possible and not just stop at one.
Thus,  we experiment with different kinds of worst-case neighboring datasets to determine which properties of datasets maximize privacy leakage for different DP-SDGs.

We begin by using small neighboring datasets with very few records.
While Nasr et al.~\cite{nasr2021adversary} use $D' = \varnothing$, when auditing deployed SDG algorithms, none of the implementations studied cover this edge case and, in fact, they even fail to generate synthetic data.
While this is a DP violation in itself, for the purposes of finding other violations that may also be present, we select neighboring datasets that are as small as possible without running into trivial runtime issues.
Thus, we select  $|D^-| = 2$, which we find to work for almost all SDG implementations.\footnote{For the DPWGAN (Synthcity) implementation, we set $|D^-| = 4$ as it fails to generate synthetic data for datasets with only two records.}
We then experiment with two different properties of the worst-case datasets, namely, ``narrow'' datasets and repeating the target record.

\descr{Narrow Datasets.}
In theory, datasets with a large number of columns (i.e., wide datasets) increase the dimensionality of the generative model and may provide the Adversary with more signals (e.g., \# of queries) that can be exploited by the MIA.
However, DP mechanisms might make each signal more noisy, thus reducing the utility of the data and making it harder to exploit the signal~\cite{ganev2023understanding}.
As we do not know how the number of columns will affect the tightness of the empirical estimates, we test our attack against narrow datasets containing only 3 columns along with the original wide datasets.

\descr{Repeating the Target Record.}
MIAs are typically evaluated against a ``fresh'' record $x_T$ that is only present in $D$ (i.e., $x_T \in D$, $x_T \notin D'$).
However, in some cases (especially in the context of anonymization),
the \textit{number of times} a target record appears in $D$ (i.e., multiplicity) can reveal sensitive information about the dataset (e.g., homogeneity attack~\cite{machanavajjhala2007diversity} against $k$-anonymity~\cite{sweeney2002k}).
Thus, we consider the setting where $x_T$ appears once in $D$ and twice in $D'$ to cover this edge case.

\section{Evaluation Framework}
In this section, we discuss our experimental setup, introducing the datasets, synthetic data generation algorithms, and the membership inference attacks we use.

\subsection{Datasets}
We experiment with two tabular datasets used to train synthetic data generation (SDG) algorithms, which have been used extensively in prior work on synthetic data~\cite{cai2021data,liu2021iterative,aydore2021differentially,mckenna2021winning} as well as in the 2018 NIST Synthetic Data Challenge~\cite{dpsynth2018}:
\begin{enumerate}
  \item \textbf{Adult}~\cite{uciadult}, used to predict whether income exceeds \$50K from Census data.
  To make sure the dataset can be used as input to all DP-SDGs, we trim and bin the dataset to 11 categorical attributes (\textit{age}, \textit{workclass}, \textit{education}, \textit{marital-status}, \textit{occupation}, \textit{relationship}, \textit{race}, \textit{gender}, \textit{hours-per-week}, \textit{native-country}, and \textit{income}). 
  \item \textbf{San Francisco Fire Dept Calls for Service (Fire)}~\cite{fire2016}, which records fire units' responses to calls made to them in 2016.
  It was used in the 2018 NIST Synthetic Data competition.
  Following prior work~\cite{annamalai2023linear}, we trim the dataset from 32 to 10 categorical attributes (\textit{ALS Unit}, \textit{Call Type Group}, \textit{Priority}, \textit{Call Type}, \textit{Zipcode of Incident}, \textit{Number of Alarms}, \textit{Battalion}, \textit{Call Final Disposition}, \textit{City} and \textit{Station Area}) to reduce the computational cost of generating thousands of synthetic datasets.%
\end{enumerate}
We focus on categorical datasets, primarily as many tabular DP-SDGs can only take those in input (e.g., PrivBayes~\cite{zhang2017privbayes}, MST~\cite{mckenna2021winning}, AIM~\cite{mckenna2022aim}, RAP~\cite{aydore2021differentially}, and GEM~\cite{liu2021iterative}) and continuous values are often binned to categorical values (up to the practitioners' discretion) to be used with these DP-SDGs.

\subsection{DP-SDG Algorithms}
In this work, we experiment with the algorithms that participated in and won the 2018 NIST Differentially Private Synthetic Data Challenge competition~\cite{dpsynth2018}.
We do so both due to their relevance and because these algorithms and their original implementations were independently verified by a team of experts to ensure the lack of privacy violations~\cite{mckenna2021winning}.
In other words, there is greater confidence that they do satisfy differential privacy. %
Yet, subtle DP violations, such as floating point bugs, have already been identified~\cite{lokna2023group}, which further motivates the need to audit these implementations. %

In particular, we focus on three of the top five submissions to the NIST competition: PrivBayes~\cite{zhang2017privbayes}, MST~\cite{mckenna2021winning}, and DPWGAN~\cite{dpwgannist}.
These have been popularly used in both research~\cite{ping2017datasynthesizer,qian2023synthcity} and industry~\cite{mstons}, and encapsulate the different ``paradigms''~\cite{annamalai2023linear} of synthetic data generation.

For MST and DPWGAN, we audit the original implementations used in the NIST competition, which are publicly available on GitHub~\cite{nistgit}, and refer to them as \textbf{MST (NIST)} and \textbf{DPWGAN (NIST)}, respectively.\footnote{In the rest of the paper, we use the Algorithm (Implementation) notation to denote the algorithm and its corresponding implementation we audit.}
However, as PrivBayes is not included in the repository, we audit the popular publicly available implementation from the DataSynthesizer repository~\cite{ping2017datasynthesizer}, which has been extensively used in prior work~\cite{stadler2022synthetic,annamalai2023linear}, and refer to this as \textbf{PrivBayes (DS)}.

Besides the three original implementations, we also audit three newer re-implementations.
Given the algorithms' popularity, many companies and research labs have since included (and potentially modified) the algorithms in their software suites.
However, as these modifications have not been independently verified, they may contain mistakes and privacy violations, once again prompting the need to audit them.
More precisely,%
we audit
\textbf{PrivBayes (Hazy)}~\cite{mahiou2022dpart}, \textbf{MST (Smartnoise)}~\cite{smartnoise}, and \textbf{DPWGAN (Synthcity)}~\cite{qian2023synthcity}.

A summary of all the algorithms tested %
and auditing methods used is reported in Table~\ref{tab:summary}.
Note that the PrivBayes implementations can only be audited using \approxdp, with $\delta = 0$, as the underlying Laplace mechanism does not satisfy \gdp.

\begin{table}[t]
  \centering
  \small
  \setlength{\tabcolsep}{5pt}
  \begin{tabular}{llll}
  \toprule
  {\bf Method} & {\bf DP } & {\bf Neighboring} & {\bf Auditing} \\
  {\bf (Implementation)} & {\bf Variant} & {\bf Dataset} & {\bf Method} \\
  \midrule
  PrivBayes (DS) & \multirow{ 2}{*}{$\varepsilon$-DP} & \multirow{ 2}{*}{Edit} & \multirow{ 2}{*}{\approxdp} \\
  PrivBayes (Hazy) &  &  &  \\
  \midrule
  MST (NIST) & \multirow{ 2}{*}{\approxdp} & \multirow{ 2}{*}{Add/Remove} & \multirow{ 2}{*}{\gdp} \\
  MST (Smartnoise) &  &  &  \\
  \midrule
  DPWGAN (NIST) & \multirow{ 2}{*}{\approxdp} & \multirow{ 2}{*}{Add/Remove} & \multirow{ 2}{*}{\gdp} \\
  DPWGAN (Synthcity) &  &  &  \\
  \bottomrule
  \end{tabular}
  \caption{Algorithms (and implementations) audited.}\label{tab:summary}
\end{table}

\subsection{MIA Instantiations}

\subsubsection{Black-Box}
\label{sec:bb_mias}
For black-box audits, we focus on two attacks widely used %
as a measure of privacy leakage from the synthetic data.

\descr{Distance to Closest Record (DCR).}
DCR is a popular heuristic used by many software libraries~\cite{houssiau2022tapas,qian2023synthcity,patki2016synthetic} and companies~\cite{tonicai,mostlyai,syntegraio,ydataai,staticeai}. %
Intuitively, synthetic data is expected to cause privacy leakage if it contains samples that are too close to the training dataset.
Formally, given synthetic data $S$ and target record $x_T$, the MIA outputs the score $-\min_{x \in S} d(x, x_T)$, for some distance metric $d$.

In our experiments, we first one-hot encode categorical features and use the Euclidean distance metric as done in prior work~\cite{houssiau2022tapas}.
Furthermore, we make the score \textit{negative} to ensure that a larger ``score'' corresponds to the presence of the target record in $D$ and is consistent with our distinguishing game that outputs $\hat{b} = 1$ if and only if $s \geq \tau$.

\descr{Querybased.}
This attack~\cite{houssiau2022tapas} builds on shadow modeling techniques.
Prior work using it include~\cite{meeus2023achilles,guepin2023synthetic,houssiau2022tapas}.
First, the adversary generates many shadow synthetic datasets from $D$ ($S_1,...,S_n$) and $D'$ ($S'_1,...,S'_n$).
Then, she evaluates the answers to queries targeted at %
$x_T$ from the shadow synthetic datasets as features.
We then train a Random Forest meta-classifier on these features to distinguish between synthetic datasets generated from $D$ and $D'$.
Finally, the adversary extracts the answers from the target synthetic dataset $S$ and returns the output of the meta-classifier on %
its features as the score.

\subsubsection{White-Box}
Unlike black-box attacks that exploit privacy leakage from the synthetic datasets, white-box attacks exploit the leakage from the trained generative model parameters directly.
As DP-SDG algorithms include a variety of generative models, ranging from simple statistical models to complex neural network architectures, the same attack cannot be generally applied to all algorithms.
Therefore, we develop and instantiate different attacks for the different DP-SDG algorithms.

\descr{PrivBayes \& MST.}
We develop a simple novel white-box attack against PrivBayes and MST that uses the shadow modeling technique.
First, the adversary generates many shadow generative models $\mathcal{G}_1,...,\mathcal{G}_n$ such that $\forall i\; \mathcal{G}_i \sim \text{GM}(D)$ and $\mathcal{G}'_1,...,\mathcal{G}'_n$ such that $\forall i\; \mathcal{G}'_i \sim \text{GM}(D')$.
Then, she extracts a set of \textit{white-box} features from each of the shadow generative models.
We experiment with two such features, namely, $\mathcal{F}_{naive}$ and $\mathcal{F}_{error}$.
In the former, the adversary simply extracts the model parameters (joint conditional probability distributions for PrivBayes and marginals for MST) directly.
In the latter, she first calculates the difference between each value in the model parameter and the corresponding exact value in $D$ and sums these differences together.
For PrivBayes and MST, each model parameter corresponds to a ``measurement'' (aka query) on the original dataset.
Intuitively, this feature set represents the total error in the ``noisy measurements'' assuming $D$ was the original dataset on which the generative model was fitted.

The adversary then trains a (Random Forest) meta-classifier on the extracted white-box features and assigns the output of the trained meta-classifier on the target generative model's extracted features as the score.
In our experiments, we find that the $\mathcal{F}_{naive}$ feature set works best for PrivBayes, while, for MST, the $\mathcal{F}_{error}$ feature set produced marginally tighter guarantees (see Appendix~\ref{app:feat_types}).
Therefore, in the rest of this work, we use $\mathcal{F}_{naive}$ for PrivBayes and $\mathcal{F}_{error}$ for MST.

\descr{DPWGAN.}
For DPWGAN, we instantiate the LOGAN attack by Hayes et al.~\cite{hayes2017logan}.
Intuitively, if the trained DPWGAN model overfits on a sensitive dataset, the discriminator will assign a higher confidence to records from that sensitive dataset.
Thus, the adversary uses the output of the trained discriminator on the target record as the score.

\subsubsection{Active White-Box}
As mentioned, we only consider the active white-box attack against %
DPWGAN.
In this setting, we instantiate the gradient canary attack by Nasr et al.~\cite{nasr2021adversary}. %
Intuitively, since the adversary can manipulate the training process of the target generative model, at each iteration, she replaces the target record's actual gradient with a \textit{canary gradient}.
Next, the gradients of each record are clipped, aggregated, and noised to satisfy DP.
The adversary calculates the dot product of the noised gradient update and the canary gradient to obtain an observation at each iteration.
Finally, she sums these observations to derive a ``score'' for the target generative model.

While Nasr et al.~\cite{nasr2021adversary} show that the active white-box attack produces tight guarantees when auditing discriminative models, we make a few modifications to the attack so that it can be applied to generative models.
Unlike discriminative models, generative ones like GANs consist of multiple models (namely, a generator and a discriminator) trained in tandem.
Since only the discriminator is trained with DP, in our work, we insert the canary gradient only in the discriminator, instead of the entire model architecture.
More precisely, we insert the Dirac canary gradient (i.e., a gradient with zeros everywhere except a single index) as this produces the tightest empirical estimates in practice~\cite{nasr2023tight}.

Also, models are often trained using software libraries like Opacus~\cite{yousefpour2021opacus} or TensorFlow Privacy~\cite{tfprivacy}, which may not readily expose the aggregated gradient for users to audit.
Therefore, we extract the gradients by calculating the difference in model parameters before and after a single iteration of training.
Although this might lead to additional terms contributing to the gradient update (e.g., the RMSProp optimizer adds a moving average to the gradient update), from a software auditing point of view, we find this method more practical to implement, and it also remains effective in producing tight empirical guarantees.
We illustrate the gradient canary attack~\cite{nasr2021adversary,nasr2023tight} with our adaptations to the DPWGAN model in Algorithm~\ref{alg:dpwgan} highlighting the changes made by the adversary to the training algorithm in red (e.g., in lines 10 to 12, the adversary replaces the gradient of the target record with a canary gradient).

\begin{algorithm}[!t]
\small
  \caption{Active white-box auditing of DPWGAN}\label{alg:dpwgan}
  \begin{algorithmic}[1]
    \Require Target record, $x_T$. \textcolor{red}{Canary gradient, $g'$}. Learning rate, $\alpha$. Clipping parameter, $c$. Batch size, $m$. Number of iterations of the critic per generator iteration, $n_\text{critic}$. Noise scale, $\sigma$. Group size, $L$. Gradient Norm bound, $c_p$.
    \Require $w_0$, initial critic parameters. $\theta_0$, initial generator's parameters.
    \State \textcolor{red}{$\text{score} \leftarrow 0$}
    \For{$t \in [T]$}
      \For{$i = 0,...,n_\text{critic}$}
        \State \textcolor{red}{$w_{start} \leftarrow w$}
        \State Pick a random sample $L_{t,i} = \{x^{(j)}\}^L_{j = 1} \sim P_{data}(x)$
        \State from the real data
        \State Sample $\{z^{(j)}\}^L_{j = 1} \sim p(z)$ a batch of prior samples
        \LineComment \textbf{Compute the per-example gradient}
        \State $g_w(x^{(j)}) = \nabla_w f_w(x^{(j)})$ for $x^{(j)} \in L_{t, i}$
        \State $g_w(z^{(j)}) = \nabla_w f_w(G(z^{(j)}; \theta))$ for $j \in [L]$
        \color{red}
        \If{$x_T \in L_{t, i}$}
          \State $g_w(x_T) = g'$
        \EndIf
        \color{black}
        \LineComment \textbf{Clip gradients}
        \For{$x^{(j)} \in L_{t, i}$}
          \State $\bar{g}_w(x^{(j)}) = g_w(x^{(j)}) / \max(1, \frac{||g_w(x^{(j)})||_2}{c_p})$
        \EndFor
        \For{$j \in [L]$}
          \State $\bar{g}_w(z^{(j)}) = g_w(z^{(j)}) / \max(1, \frac{||g_w(z^{(j)})||_2}{c_p})$
        \EndFor
        \LineComment \textbf{Add Noise}
        \State $\tilde{g}_w = \frac{1}{L}\left(\sum^L_{j = 1} \bar{g}_w(x^{(j)}) + \mathcal{N}(0, \sigma^2c^2_p\textbf{I})\right) - $
        \State $\frac{1}{L} \sum^L_{j = 1} \bar{g}_w(z^{(j)})$
        \State $w \leftarrow w + \alpha \cdot \text{RMSProp}(w, \tilde{g}_w)$
        \State $w \leftarrow \text{clip}(w, -c, c)$
        \State \textcolor{red}{$w_{diff} \leftarrow w - w_{start}$}
        \State \textcolor{red}{$\text{score} \leftarrow \text{score } + \langle w_{diff}, g' \rangle$}
      \EndFor
      \State Sample $\{z^{(j)}\}^L_{j = 1} \sim p(z)$ a batch of prior samples
      \State $g_\theta \leftarrow - \nabla_\theta \frac{1}{m} \sum^m_{j = 1} f_w(G(z^{(j)}; \theta))$
      \State $\theta \leftarrow \theta - \alpha \cdot \text{RMSProp}(\theta, g_\theta)$
    \EndFor
  \color{red}
  \State \Return score, $\theta$, $w$
  \color{black}
  \end{algorithmic}
\end{algorithm}

\section{Experimental Results}
This section presents our experimental evaluation geared to audit six state-of-the-art DP-SDG implementations with different, increasingly stronger threat models.
First, we compare black-box attacks and analyze the differences in empirical guarantees ($\empeps$) with \textit{average-case} and \textit{worst-case} neighboring datasets.
We then experiment with white-box attacks (namely, LOGAN~\cite{hayes2017logan} for DPWGAN and a novel attack for PrivBayes and MST) as well as an {\em active} white-box one (adapting Nasr et al.~\cite{nasr2023tight}'s attack to the generative setting).
Finally, we investigate whether our auditing procedure can identify common DP violations and discover new DP ones.

For each experiment, we train 10,000 SDG models; we use 6,000 as shadow models to train the meta-classifier for attacks that use shadow modeling (for attacks that do not, we do not train any shadow models), and 2,000 models to choose the optimal threshold yielding the largest lower bound $\empeps$.
We then test all attacks on the remaining 2,000 models and calculate the false positive and false negative rates needed for the $\empeps$ estimation.
Following prior work~\cite{nasr2021adversary,nasr2023tight}, all lower bounds are given with 95\% confidence (Clopper-Pearson~\cite{clopper1934use}).
We also report error bars, which we obtain via five-fold cross-validation -- i.e., we split the 10,000 models into 5 partitions of 2,000 models each and repeatedly test the attack on each of the five partitions using the other four partitions to train the meta-classifier and choose the optimal threshold.

The source code needed to reproduce our experiments is available from \url{https://github.com/spalabucr/synth-audit}.

\subsection{Black-Box Auditing}\label{sec:bb}

\subsubsection{Average-Case Dataset}

\begin{figure}[t]
  \centering
  \begin{subfigure}[b]{\mywidth\linewidth}
    \includegraphics[width=\mywidth\linewidth]{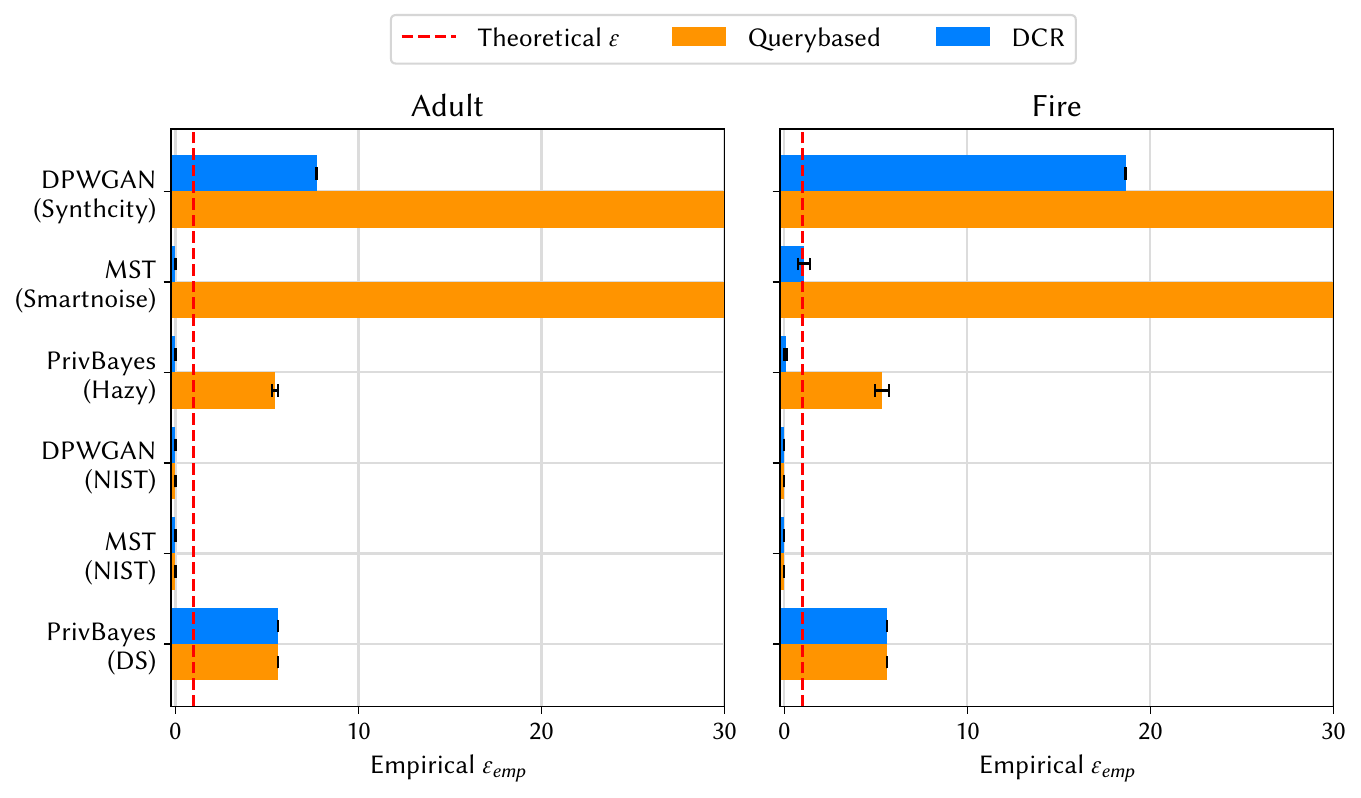}
    \caption{Theoretical $\varepsilon = 1.0$}
  \end{subfigure}
   \begin{subfigure}[b]{\mywidth\linewidth}
    \includegraphics[width=\mywidth\linewidth]{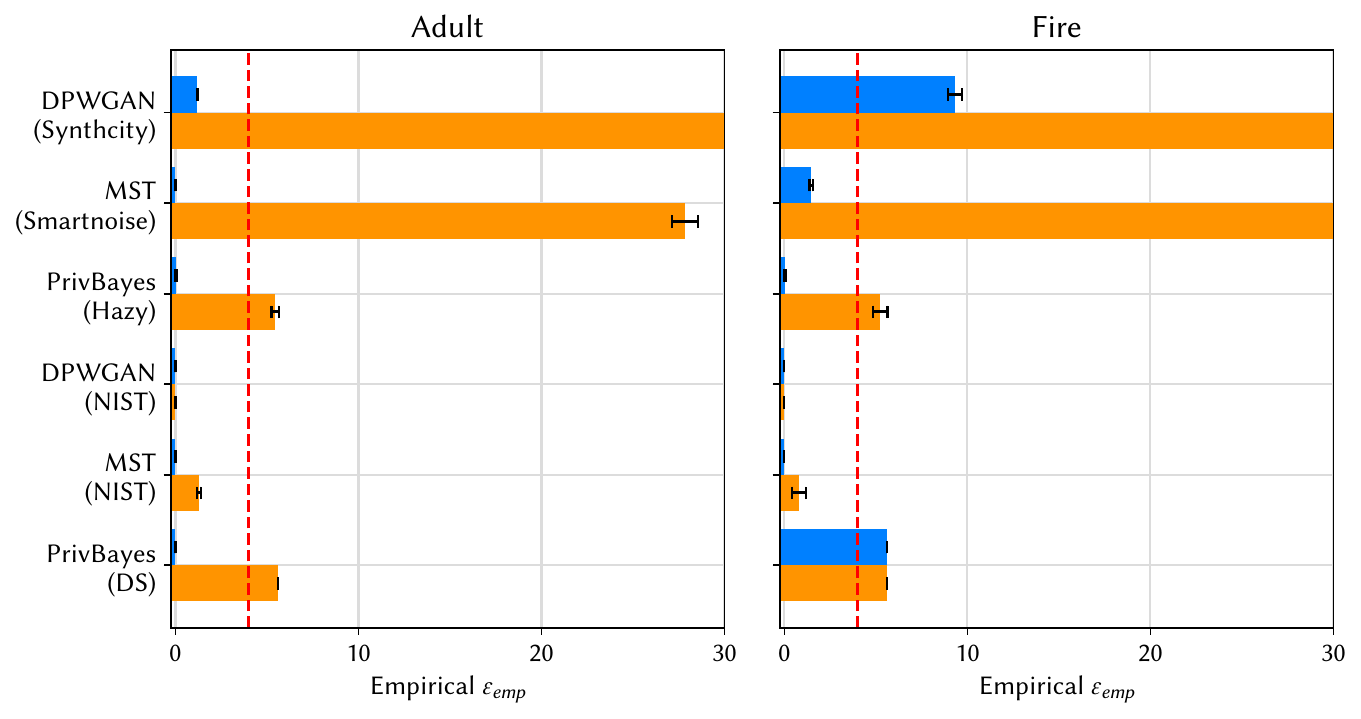}
    \caption{Theoretical $\varepsilon = 4.0$}
  \end{subfigure}
  \caption{Black-box auditing, Querybased and DCR attacks.}
  \label{fig:qb_vs_dcr}
\end{figure}

\begin{figure}[t]
  \centering
  \begin{subfigure}[b]{\mywidth\linewidth}
    \includegraphics[width=\mywidth\linewidth]{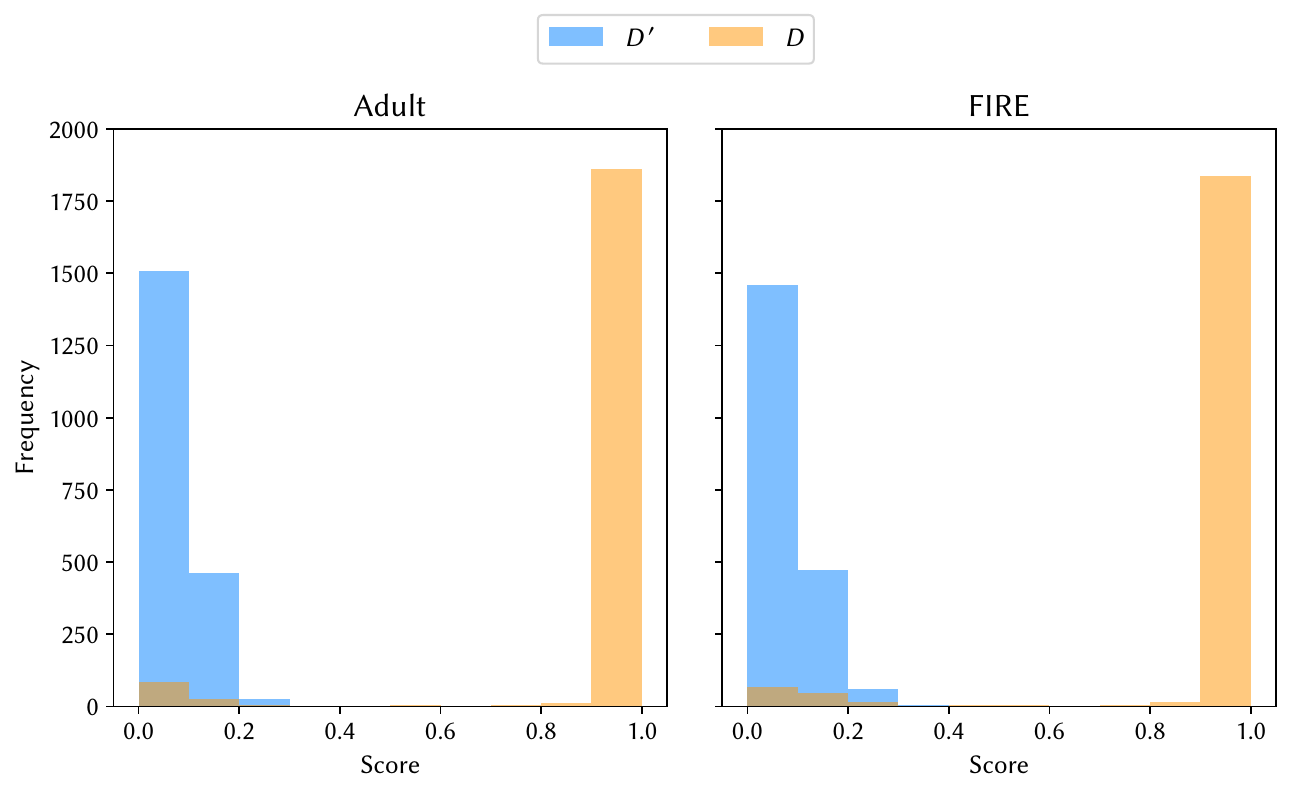}
    \caption{Querybased}
  \end{subfigure}
   \begin{subfigure}[b]{\mywidth\linewidth}
    \includegraphics[width=\mywidth\linewidth]{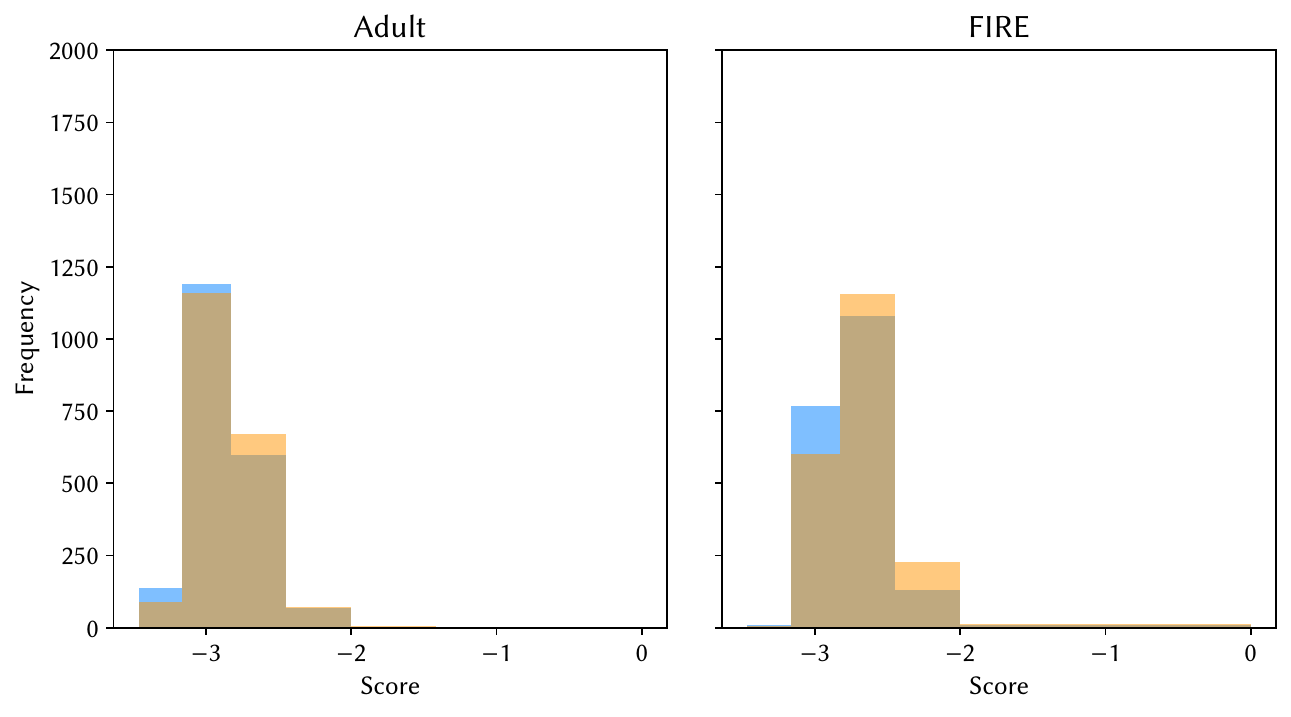}
    \caption{DCR}
  \end{subfigure}
  \caption{Distribution of Querybased/DCR attack scores against PrivBayes (Hazy) trained on $D$ vs $D'$ at $\varepsilon = 1.0$.}
  \label{fig:qb_vs_dcr_scores}
\end{figure}

We start by auditing the DP-SDG implementations by instantiating the DP distinguishing game (Section~\ref{sec:dp_game}) with an average-case dataset ($|D| = 1000$) and two popular black-box attacks, i.e., Querybased and DCR~\cite{houssiau2022tapas}.
In Figure~\ref{fig:qb_vs_dcr}, we report the empirical $\empeps$ guarantees for the six DP-SDG implementations, using Adult and Fire datasets, at two $\varepsilon$ values corresponding to high and moderate privacy -- respectively, $\varepsilon = 1.0$ and $\varepsilon = 4.0$.

\descr{DP Violations.} We observe that the empirical privacy leakage is much larger than the theoretical guarantee (i.e., $\empeps \gg \varepsilon$) for {\em all} the implementations not submitted to the NIST competition, i.e., DPWGAN (Synthcity), MST (Smartnoise), PrivBayes (Hazy), and PrivBayes (DS).
By manually inspecting the code, we find that these directly extract the metadata from the input dataset. %
This ``metadata violation'' occurs as metadata like categories, minimum/maximum numerical value, etc., might unexpectedly leak information, especially for vulnerable target records with rare values~\cite{stadler2022synthetic}.
Although this violation was already identified in 2022 for numerical datasets in PrivBayes (DS) and PATE-GAN~\cite{jordon2018pate}, it still remains unfixed in the DataSynthesizer library.
The same infringement also occurs in other implementations, such as PrivBayes (Hazy), MST (Smartnoise), and DPWGAN (Synthcity).
Note that we reported these and all other violations to the respective library authors; see Section~\ref{sec:ethics}.

While Querybased identifies violations in 16 out of 24 experiments, DCR only identifies 7 of them. %
For the 9 violations identified by Querybased but not by DCR, the AUC of the latter is close to random ($\approx 0.5$), while that of the former is $\geq 0.95$.
In other words, DCR not only misses privacy violations but also severely underestimates privacy leakage from synthetic data. %
Evidently, it is ineffective at providing an effective measure of privacy and, in practice, should not be used to evaluate (differentially private) synthetic data.

\descr{Querybased vs DCR.} We then investigate \textit{why} this may be happening. %
To do so, we plot the raw scores output by the DCR and Querybased attacks against the PrivBayes (Hazy) implementation in Figure~\ref{fig:qb_vs_dcr_scores}.
Recall from Section~\ref{sec:dp_game} that these represent the confidence the attack assigns to SDG being trained on $D$; specifically, Querybased outputs a \textit{probability} score (i.e., $s \in [0, 1]$), whereas DCR outputs a \textit{distance}, which we make negative (i.e., $s \in (-\infty, 0]$), as discussed in Section~\ref{sec:bb_mias}.
Regardless, for both attacks, higher scores represent stronger confidence.

For Querybased, the distinct score separation when the DP-SDG is fitted on $D$ and $D'$ indicates that the meta-classifier learns and exploits the queries targeted at the target record effectively.
For DCR, the distances between the target record and the closest synthetic record remain relatively similar regardless of whether the DP-SDG was fit on $D$ or $D'$.
As DCR relies on whole target records being memorized and output by the SDG, it seems unable to exploit more complex ways in which information can be leaked from synthetic data~\cite{annamalai2023linear}.

\descr{Loose Estimates.} Finally, auditing using black-box attacks results in several empirical privacy estimates much smaller than the theoretical upper bounds (i.e., $\empeps \ll \varepsilon$).
Interestingly, this happens for the DP-SDG implementations submitted to NIST, as $\empeps \approx 0$ for MST (NIST) and DPWGAN (NIST), with both $\varepsilon = 1.0$ and $4.0$.
Arguably, it is unclear whether this is due to 1) leakage not being maximized under average-case neighboring datasets or 2) state-of-the-art black-box attacks being limited in power.
To answer this, we next evaluate black-box attacks using \textit{worst-case} neighboring datasets.

\subsubsection{Worst-Case Datasets}
\begin{figure}[t]
  \centering
  \includegraphics[width=\mywidth\linewidth]{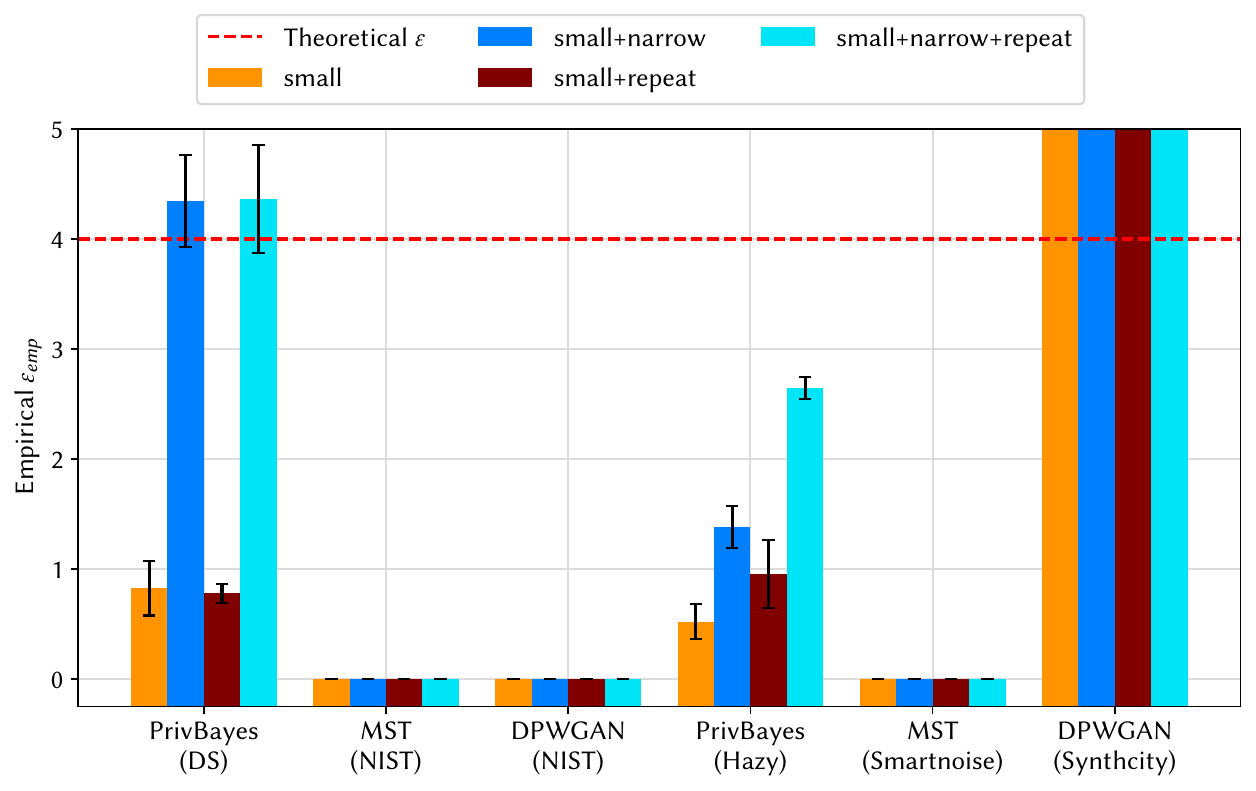}
  \caption{Black-box auditing at $\varepsilon = 4.0$ with different {\em worst-case} datasets using the Querybased attack.}
  \label{fig:compare_worstcase_datasets}
\end{figure}

To avoid the metadata violation discussed above, we craft worst-case neighboring datasets such that the domain of $D$ is the same as $D'$ (thus, the metadata extracted will be the same).
Additionally, as the exact worst-case dataset could potentially be algorithm-dependent, we use worst-case datasets with a small number of rows (\texttt{small}) and experiment with two properties, i.e., 1) a small number of columns (\texttt{narrow}) and 2) repeating the target record (\texttt{repeat}).
Finally, to focus on auditing the underlying noise-addition mechanism of the implementations, we use a provisional dataset to standardize the ``structure'' of the DP-SDGs, as done by the top NIST submissions~\cite{mckenna2021winning}.
More precisely, in the rest of the experiments, we standardize the Bayesian network built by PrivBayes and the marginals selected by MST across all models.
In Figure~\ref{fig:compare_worstcase_datasets}, we plot the empirical guarantees ($\empeps$) obtained by auditing using Querybased, for different worst-case datasets at theoretical $\varepsilon = 4.0$.

\descr{DP Violation.} We find a DP violation for DPWGAN (Synthcity), regardless of the type of worst-case dataset.
Recall that we prevent metadata violations by design (see Section~\ref{sec:bb}); thus, this must be caused by something else.
After manually inspecting the source code, we found that a random seed was re-used by the library for reproducibility, but this was not mentioned in the sample code.
This removes any randomization from the code, thus making it deterministic, which results in a major DP violation.
Alas, this type of bug was also found in other DP libraries, e.g., JAX~\cite{prngreuse}.

\descr{Implementation-dependent worst-case.} Next, we find that, even for the same algorithm, the worst-case dataset can be specific to the \textit{implementation}.
For PrivBayes (Hazy), the \texttt{small+narrow+repeat} dataset yields the highest privacy leakage estimate ($\empeps = 2.64$), much higher than the estimate for \texttt{small+narrow} ($\empeps = 1.38$).
Whereas for PrivBayes (DS), the estimates are roughly the same for \texttt{small+narrow+repeat} and \texttt{small+narrow} datasets ($\empeps = 4.36$ and 4.35, respectively).
Ostensibly, this is due to each implementation introducing specific additional steps (pre-processing, validation, etc.), which might subtly alter the overall privacy leakage of the implementation itself.

\descr{Zero $\empeps$.} Finally, for MST (NIST), MST (Smartnoise), and DPWGAN (NIST), we see that $\empeps \approx 0$ for all worst-case datasets.
This suggests that even in the worst-case setting, state-of-the-art black-box MIAs are not powerful enough to exploit the privacy leakage from these DP-SDG implementations.
DP-SDGs' theoretical guarantees typically apply to the underlying generative models directly and only transfer to the generated synthetic data through the post-processing theorem of DP;
thus, loose estimates could also be due to the inability of state-of-the-art black-box MIAs to fully exploit the information available from DP-SDGs, which motivates us to explore stronger threat models.

\begin{figure}[t]
  \centering
  \includegraphics[width=0.9\linewidth]{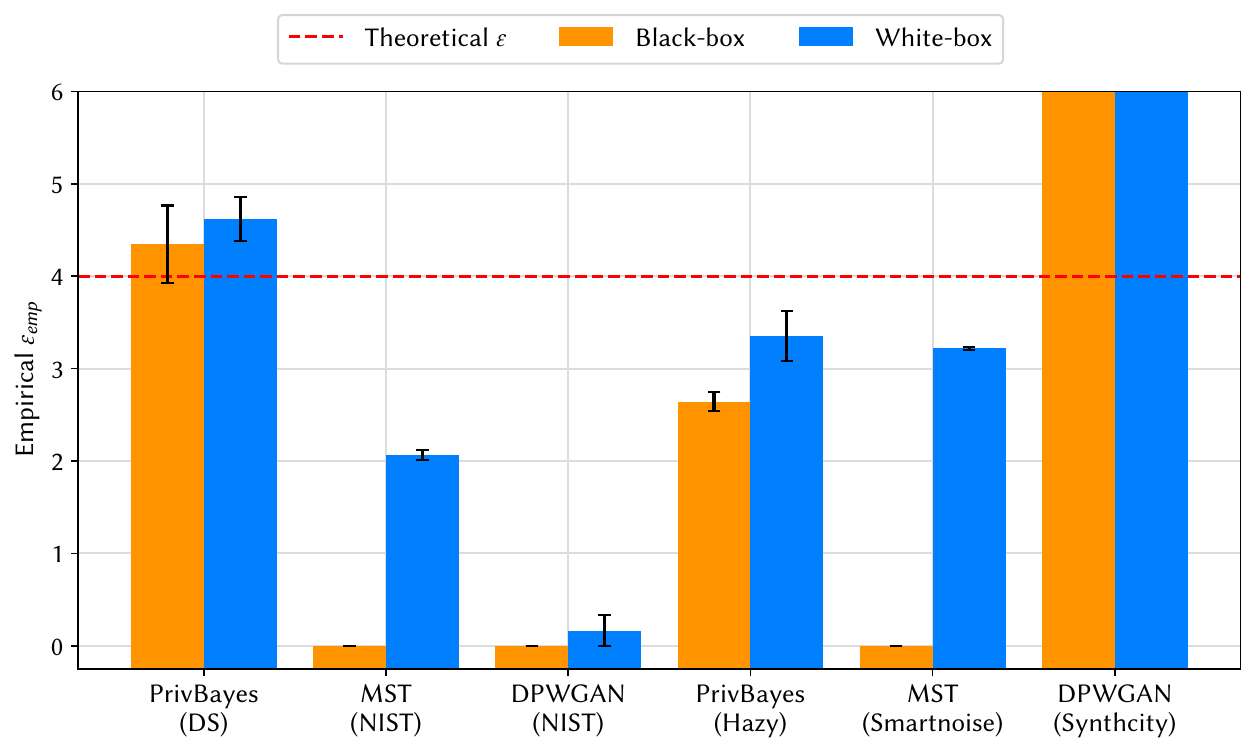}
  \caption{White-box vs black-box auditing at $\varepsilon = 4.0$ using implementation-specific worst-case neighboring datasets.}
  \label{fig:bb_vs_wb_worstcase}
\end{figure}

\subsection{White-Box Auditing}
We now move on to adversaries with stronger capabilities, i.e., with access to the (final) fitted generative model.
Specifically, we use the LOGAN~\cite{hayes2017logan} white-box attack against DPWGAN and a novel one against PrivBayes and MST.

First, we determine if white-box attacks can exploit the additional information available compared to black-box attacks.
Specifically, we compare the $\empeps$ values obtained with the white-box attacks vs.~the Querybased black-box one in Figure~\ref{fig:bb_vs_wb_worstcase}.
Experiments consider DP-SDGs trained with $\varepsilon = 4.0$ on the worst-case pair of neighboring datasets that produces the largest empirical guarantees for each DP-SDG implementation (which we find empirically, as discussed in Appendix~\ref{app:compare_worstcase_wb}).

\descr{White- vs Black-box.} For almost all implementations, the white-box attacks result in significantly tighter $\empeps$ estimates, i.e., closer to the theoretical $\varepsilon$-s, than the black-box attack.
This is particularly evident for MST (NIST) and MST (Smartnoise), where the former obtain $\empeps = 2.06$ and $3.22$, respectively, while the latter is unable to detect any leakage (i.e., $\empeps \approx 0$).
Similarly, for PrivBayes (DS) and PrivBayes (Hazy), white-box audits produce tighter estimates of $4.62$ and $3.35$, respectively, compared to $4.35$ and $2.64$ for black-box attacks. %

Note that, in the white-box setting, $\empeps > \varepsilon$ for PrivBayes (DS) indicates another DP violation in the DataSynthesizer library.
This violation was not obvious in the black-box setting as the standard deviation of $\empeps$ was larger, whereas, in the white-box setting, the theoretical $\varepsilon$ is well outside the standard deviation of $\empeps$.
As mentioned previously, PrivBayes (DS) includes a number of pre-processing steps, other than inferring the metadata, that may not satisfy DP.
Thus, we believe this is the most likely cause of the DP violation here.

\begin{figure}[t]
  \centering
  \includegraphics[width=\mywidth\linewidth]{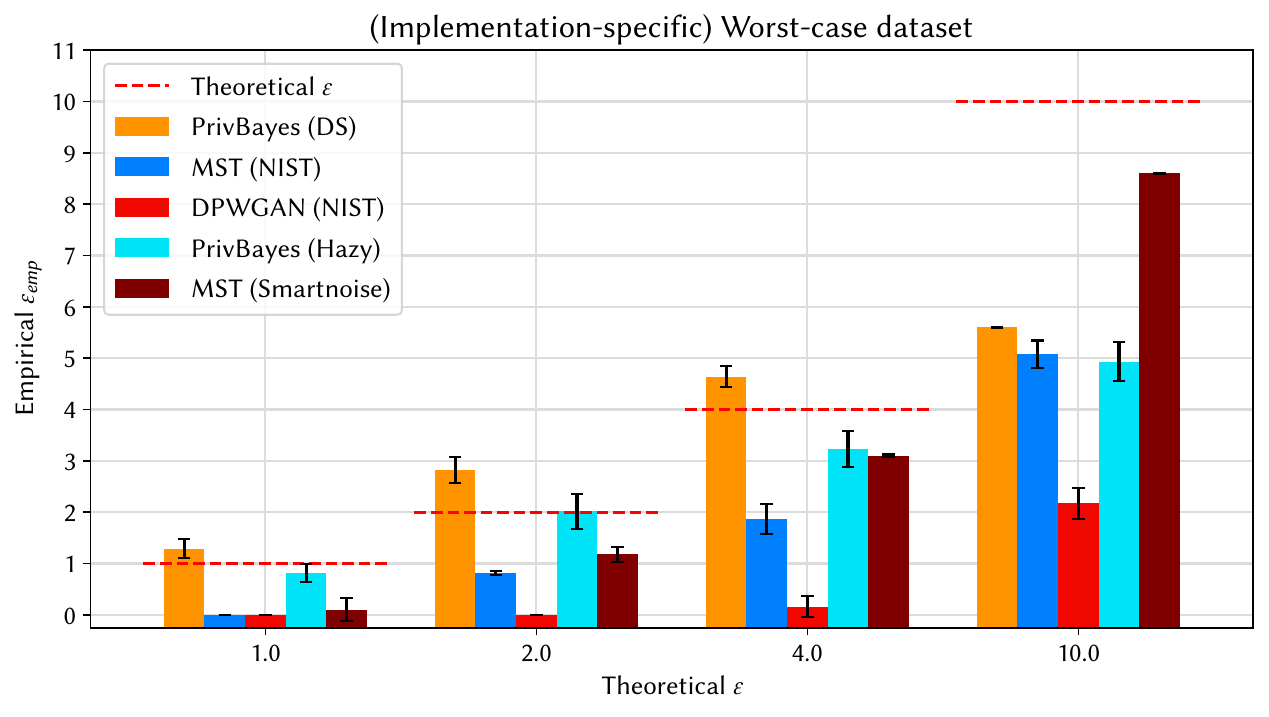}
  \caption{White-box auditing at $\varepsilon = 1.0, 2.0, 4.0, 10.0$.}
  \label{fig:worstcase_audit_multeps}
\end{figure}

\descr{Impact of privacy parameter.} Next, as DP-SDGs can be instantiated with different levels of privacy (typically depending on the use case), we investigate whether our white-box audits can produce tight estimates at different $\varepsilon$ values;
see Figure~\ref{fig:worstcase_audit_multeps}.
The empirical estimates are tightest for PrivBayes (DS) and PrivBayes (Hazy) across most $\varepsilon$-s.\footnote{Note that we do not consider PrivBayes (Hazy) at $\varepsilon = 2.0$ a privacy violation as it lies within the standard deviation of $\empeps = 2.01 \pm 0.35$.}
Furthermore, $\empeps$ values for MST (NIST) and MST (Smartnoise) grow consistently with increasing $\varepsilon$, which indicates that the white-box attacks do leverage the increasing privacy leakage in practice.
However, the estimates are not as tight as those of the PrivBayes implementations, especially at smaller $\varepsilon$-s.
This may be due to the domain compression techniques used by MST; these are more aggressively applied at smaller $\varepsilon$-s, and might result in a loss of information, thus making it harder for white-box attacks to precisely estimate the privacy leakage.
Nevertheless, using our white-box attacks, %
and worst-case dataset setting, we can audit the MST implementations much more tightly than prior work~\cite{houssiau2022tapas}.

Finally, we find that white-box auditing does not produce tight $\empeps$ estimates for DPWGAN (NIST), even in the worst-case neighboring dataset setting.
This indicates that not even the white-box adversary is powerful enough in this setting, thus motivating us to consider {\em active} white-box attacks. %

\subsection{Active White-Box Auditing}
\begin{figure}[t]
  \centering
  \includegraphics[width=\mywidth\linewidth]{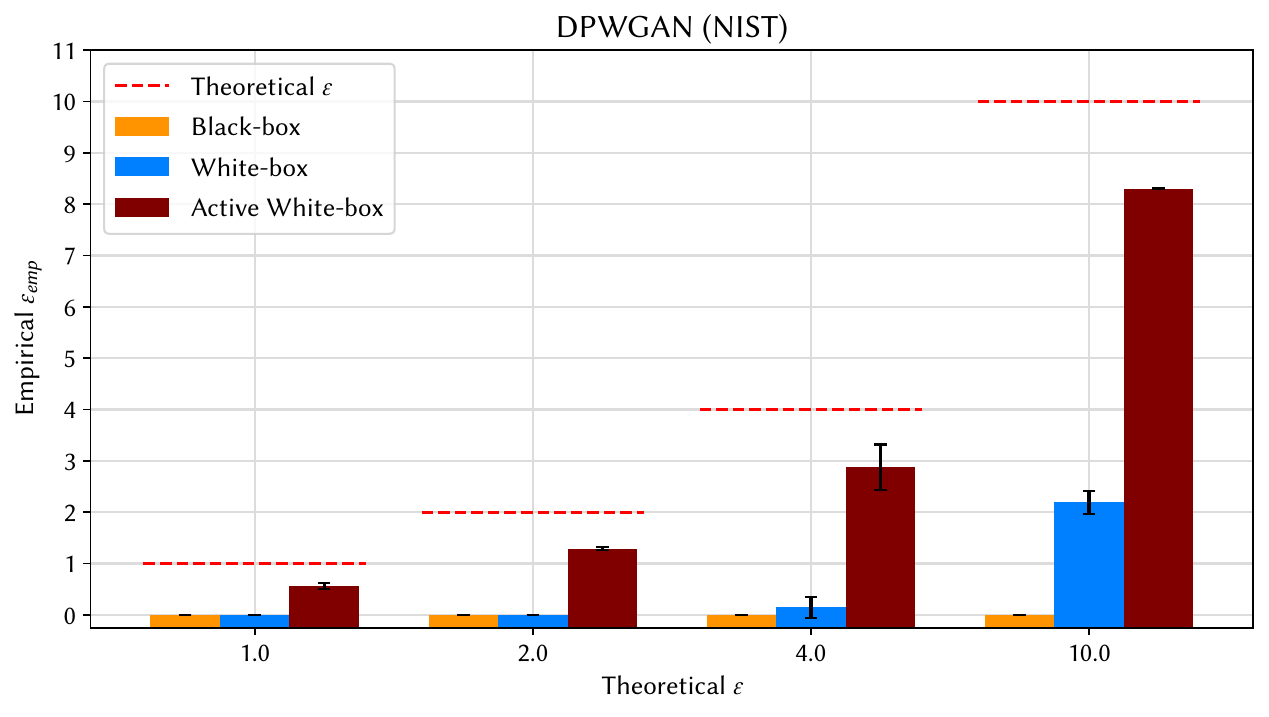}
  \caption{Black-box vs white-box vs active white-box auditing DPWGAN (NIST) at $\varepsilon = 1.0, 2.0, 4.0, 10.0$.}
  \label{fig:wb_vs_active_wb}
\end{figure}

As discussed in Section~\ref{sec:threat}, in the active white-box attack, the adversary manipulates training by inserting arbitrary gradients into the model, in this case, DPWGAN's discriminator.

In Figure~\ref{fig:wb_vs_active_wb}, we report the resulting $\empeps$ estimates, using a worst-case neighboring dataset (\texttt{small+repeat}), to audit DPWGAN (NIST) at various theoretical $\varepsilon$-s .
For completeness, we also report the $\empeps$ values for the black-box (Querybased) and white-box (LOGAN) attacks.
We observe that the active attack produces relatively tight empirical $\empeps$ estimates, especially with large $\varepsilon$-s.
Specifically, for $\varepsilon = 1.0, 2.0, 4.0, 10.0$, we obtain, respectively, $\empeps = 0.56, 1.29, 2.88, 8.31$.
This confirms that for DPWGAN, unlike PrivBayes and MST, auditing using the strongest active white-box attack is necessary to produce tight empirical estimates.

\subsection{Finding Other DP Violations}

The experimental analysis presented above allows us to identify the threat models and adversarial capabilities needed to tightly audit different DP-SDG implementations, highlighting the prevalence of metadata violations in DP-SDGs. %
Nonetheless, there could also be more subtle/less egregious violations.
These are inherently harder to identify, and previous work had to rely on manual code inspection by experts to verify DP guarantees and find DP violations in DP-SDG implementations~\cite{dpsynth2018}.

In the rest of this section, we investigate whether our auditing procedure can identify DP violations and verify DP guarantees \textit{automatically}.
In fact, using our auditing procedure, we identify a new violation in the implementation of DPWGAN submitted to NIST, along with violations that were previously found through manual inspection.

\subsubsection{Early Stopping}

\begin{figure}[t]
  \centering
  \includegraphics[width=\mywidth\linewidth]{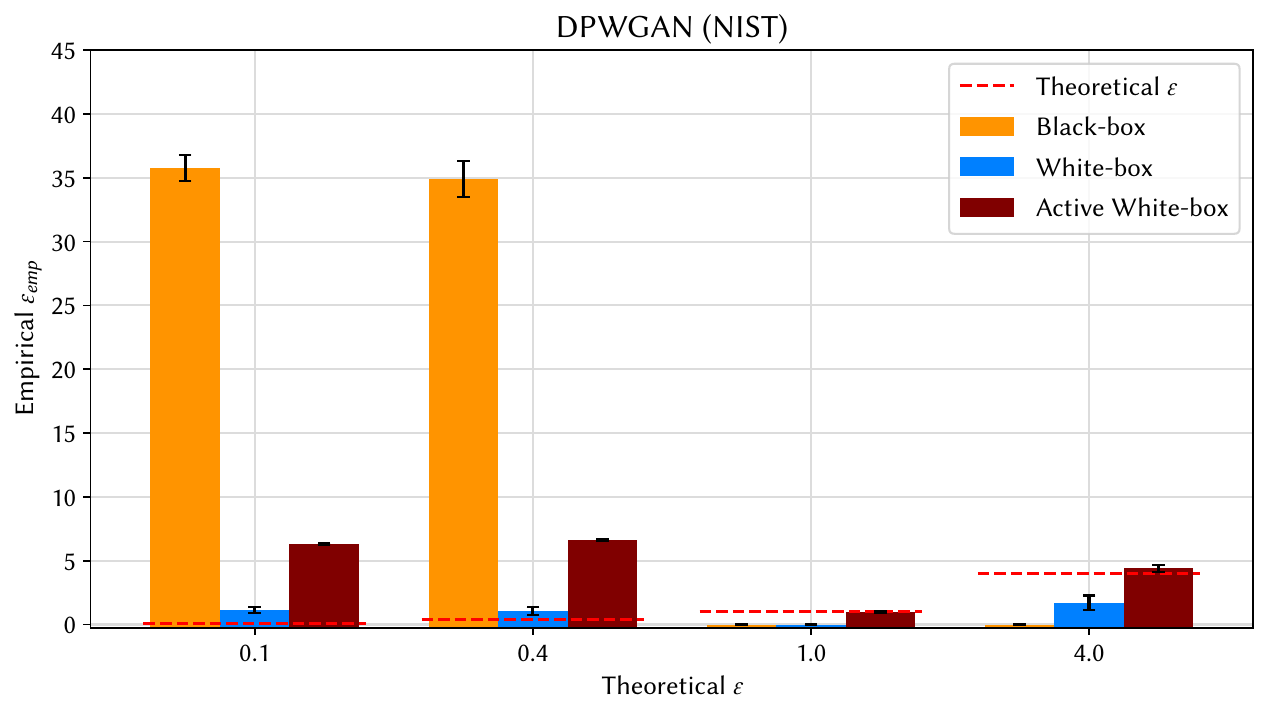}
  \caption{Black-box auditing of DPWGAN.} %
  \label{fig:dpwgan_bug}
\end{figure}

Recall that the DPWGAN (NIST) implementation was submitted to the Differentially Private Synthetic Data Challenge competition~\cite{dpsynth2018};
participants had to submit their code and a technical report %
proving their algorithm satisfies DP~\cite{mckenna2021winning}.
The experts that reviewed both did not identify any violations, in other words, confirming that the implementation does satisfy \approxdp. %
Our experiments presented so far also support this.

However, as seen earlier, tight empirical estimates may only be possible in \textit{worst-case} settings.
DPWGAN is a much more complex algorithm than PrivBayes/MST, involving many hyper-parameters that can be tuned.
While we have only looked at worst-case target record and worst-case neighboring datasets thus far, for DPWGAN, we now look at worst-case \textit{hyper-parameters} as well.
After experimenting with different worst-case hyper-parameters, we find a DP violation when the \texttt{batch\_size} hyperparameter is set to 1.
In Figure~\ref{fig:dpwgan_bug}, we plot the $\empeps$ estimates with \texttt{batch\_size} set to 1, %
finding that, with small $\varepsilon $ values ($0.1, 0.4$), $\empeps \gg \varepsilon$. %
Specifically, auditing using the black-box attack with $\varepsilon = 0.1$ and $0.4$ results in empirical estimates $\empeps = 35.8$ and $34.9$, respectively, while $\empeps < \varepsilon$ for $\varepsilon = 1.0$ and $4.0$.

We believe that the issue stems from the ``early stopping'' feature of the privacy accounting method.
In the code, a privacy accountant tracks the privacy budget at each iteration, and training is aborted when that is exceeded.
DPWGAN (NIST) applies two different accountants, depending on $\varepsilon$: %
for $\varepsilon < 0.7$, the privacy accountant is data-dependent as it uses the size of the dataset ($|D|$) without adding differentially private noise.
Therefore, the model goes through a different number of iterations when trained on $D$ and $D'$, which is exploited by the black-box attack.
Interestingly, the white-box and active white-box attacks do not detect a large DP violation in this setting since they were only auditing the \textit{discriminator}, whereas this bug affects the \textit{generator} more significantly.
As is not clear to us how to effectively attack the generator in the white-box/active white-box settings, we believe this prompts an interesting area for further research.

\label{sec:noise_scale_bug}
\begin{figure}[t]
  \centering
  \includegraphics[width=\mywidth\linewidth]{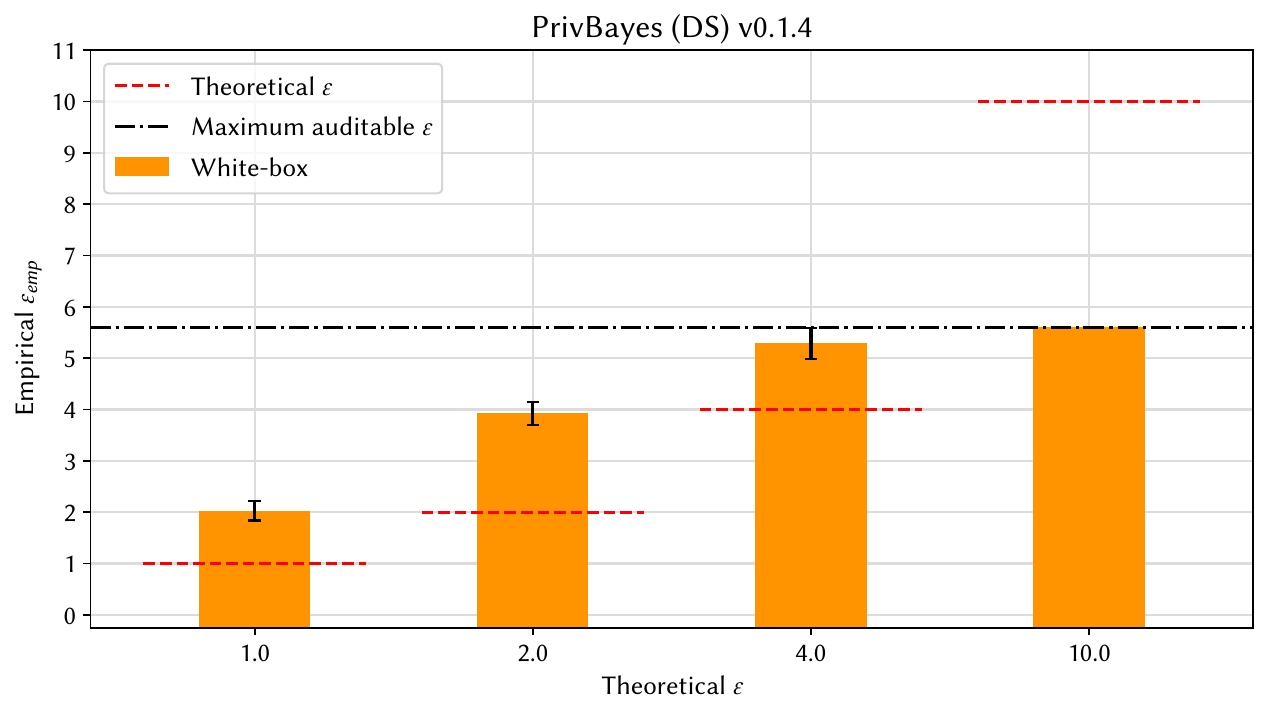}
  \caption{White-box auditing of DataSynthesizer v0.1.4.} %
  \label{fig:ds_0.1.4}
\end{figure}

\subsubsection{Noise Scale}

One issue with DP-SDG implementations previously found through manual inspections is the noise scale bug, where the noise parameter of the algorithm is incorrectly configured.
For instance, this was identified in the PrivBayes (DS) implementation at v0.1.4 and raised as a GitHub issue by a privacy researcher.\footnote{\url{https://github.com/DataResponsibly/DataSynthesizer/issues/34}}
Unlike the early stopping bug, this one does not completely break the DP guarantees of the algorithm; rather, it results in the DP guarantee (slightly) overestimating the actual privacy protections. %
Thus, this type of bug can only be caught if the auditing procedure is tight (i.e., $\empeps \approx \varepsilon$);
otherwise, even if the actual $\varepsilon$ parameter is much larger than what is claimed, %
the auditing procedure may not return $\empeps \gg \varepsilon$, thus not flagging it as a DP violation.

Next, we investigate whether our auditing procedure can automatically detect the noise scale bug without the need for manual expert analysis.
Specifically, we use our white-box attack to audit PrivBayes (DS) v0.1.4.
Figure~\ref{fig:ds_0.1.4} shows the results of our white-box auditing procedure at various levels of theoretical $\varepsilon$.
For $\varepsilon = 1.0$ and $2.0$, our auditing procedure produces empirical estimates $\empeps = 2.03$ and $3.92$, respectively---clearly flagging the DP violation.
It actually calculates the \textit{magnitude} of the violation accurately, as the \textit{true} $\varepsilon$ is approximately twice the {\em claimed}  $\varepsilon$.

However, at $\varepsilon = 4.0$, our procedure only identifies that the noise scale is configured wrongly but does not accurately calculate the magnitude of the error; at $\varepsilon = 10.0$, it does not even detect the violation as it reaches the \textit{maximum auditable $\varepsilon$} limit.
Recall from Section~\ref{sec:prelims_audit_dp} that the maximum auditable $\varepsilon$ limit is an inherent limitation of our auditing procedure.

\subsection{Takeaways}
Our experimental analysis shows that the state-of-the-art \textit{black-box} MIAs commonly used to evaluate privacy in DP-SDGs may be severely limited in power.
For instance, the DCR heuristic vastly underestimates privacy leakage from the synthetic data, often achieving AUCs close to a random guess ($\approx 0.5$), in settings where Querybased yields $\geq 0.95$ AUC.
With the latter, we also find metadata violations in many DP-SDG implementations (stemming from learning metadata directly from the input dataset).
Nonetheless, auditing using Querybased still generally results in loose empirical estimates of privacy leakage, even in worst-case settings.

Arguably, to estimate the privacy leakage from DP-SDGs tightly, we need implementation-specific worst-case datasets and stronger threat models.
Our experiments show that the privacy leakage of different DP-SDG implementations is maximized for different types of worst-case datasets.
Auditing using a novel \textit{white-box} attack yields tight estimates for PrivBayes and MST implementations; however, for machine learning models like DPWGAN, we need attacks in a much more powerful, \textit{active white-box} threat model.
When auditing PrivBayes (Hazy), MST (Smartnoise), and DPWGAN (NIST) at $\varepsilon = 4.0$ using our auditing procedure, we obtain nearly tight empirical privacy estimates of $\empeps = 3.23$, $3.10$, and $3.02$, respectively.
In comparison, using %
Querybased only achieves loose estimates of $\empeps = 2.64$, $0.00$, and $0.00$, respectively.

Last but not least, our auditing procedure can find several DP violations (e.g., noise scale bug) in DP-SDGs \textit{automatically}, without the need for manual inspection.
It also identifies a new DP violation in the DPWGAN (NIST) implementation.

\section{Related Work}

\noindent\textbf{DP-SDG Auditing.} To our knowledge, our work presents the first large-scale audit of DP-SDG algorithms and implementations.
Houssiau et al.~\cite{houssiau2022tapas} introduce, and audit DP-SDGs with, the TAPAS toolbox, though for only a single implementation of a single DP-SDG.
They empirically estimate the privacy guarantees of MST~\cite{mckenna2021winning} at $\varepsilon = 10$ using the black-box Querybased attack, but only achieve loose guarantees.
By contrast, our audits %
include multiple algorithms/implementations, stronger threat models, %
and are considerably tighter. %

\descr{MIAs against synthetic data.}
The DCR heuristic is one of the earliest black-box methods used for MIAs~\cite{hilprecht2019monte,lu2019empirical}, although having limited effectiveness.
Stadler et al.~\cite{stadler2022synthetic} use shadow modeling to show that outlier records in synthetic data are often vulnerable to black-box MIAs.
They also find DP metadata violations in implementations of PrivBayes and PATE-GAN (i.e., they extract metadata from the input dataset).
However, their goal meaningfully differs from ours as they do not focus on auditing and only consider black-box attacks.

The only white-box MIA against SDGs is LOGAN~\cite{hayes2017logan}, which attacks GANs (we use it for DPWGAN); we present the first white-box MIAs against PrivBayes~\cite{zhang2017privbayes} and MST~\cite{mckenna2021winning}.
\descr{Empirically Estimating Privacy in DP-ML.} %
Prior work has extensively focused on empirically estimating the privacy of differentially private discriminative models (DP-ML) in both centralized and federated settings~\cite{wei2023dpmlbench,nasr2021adversary,nasr2023tight,zanella2023bayesian,galen2024oneshot,maddock2022canife,jayaraman2019evaluating,steinke2023privacy,jagielski2020auditing,kulynych2019disparate}.
Jayaraman et al.~\cite{jayaraman2019evaluating} and Jagielski et al.~\cite{jagielski2020auditing} present auditing schemes for DP-ML but generally only achieve loose empirical estimates.
Nasr et al.~\cite{nasr2021adversary} audits DP-ML using the \approxdp definition and Clopper-Pearson intervals, requiring a million runs at $\varepsilon = 10$.
Zanella-Béguelin et al.~\cite{zanella2023bayesian} %
focus on reducing the number of training runs using so-called {\em credible} intervals, while Nasr et al.~\cite{nasr2023tight} audit DP-ML using the \gdp definition and credible intervals and show that 1,000 runs are in fact enough to audit models at $\varepsilon = 10$.
While auditing with 1,000 runs is generally feasible for centralized learning, it might be less so in resource-constrained settings typical of federated learning; thus, another line of work focuses on reducing the number of runs to one~\cite{maddock2022canife,galen2024oneshot,steinke2023privacy}.

\descr{Tightly Auditing DP-ML.} Nasr et al.~\cite{nasr2023tight} present a tight auditing scheme for discriminative models trained using differentially private stochastic gradient descent~\cite{abadi2016deep}.
They show that {\em natural} (i.e., not adversarially crafted) datasets are enough for tightness, %
considering a white-box adversary who can choose arbitrary {\em canary} gradients at each step.
\longVer{Inspired by their work, we also go beyond black-box adversaries, introducing novel white-box MIAs and adapting their canary-based (active) attack to generative models.}
Arguably, our work is broader in nature and scope.
We consider three different training algorithms for generative models, compared to just stochastic gradient descent.
Also, in discriminative models, there is a single signal (i.e., the model's loss on the target record) that can be exploited by MIAs, whereas  generative models' outputs lie in a higher dimensional space, producing many possible signals. %
Thus, we experiment with multiple MIAs for each threat model %
(e.g., Querybased and DCR for black-box), 
studying the disparity of their effectiveness. %
Incidentally, we find that in DP-SDGs, unlike discriminative models, adversarially crafted implementation-specific worst-case datasets are necessary to achieve tightness. %

\descr{Auditing DP Implementations.}
Prior audits of DP implementations include DP-Sniper~\cite{bichsel2021dp}, DP-Opt~\cite{niu2022dp}, and Delta-Siege~\cite{lokna2023group}.
Note that~\cite{bichsel2021dp,niu2022dp} do not consider DP-SDGs, while~\cite{lokna2023group} is orthogonal to our work as it aims to amplify existing distinguishers and classifiers to identify floating-point DP violations
in DP implementations (including MST~\cite{mckenna2021winning}).

\longVer{Overall, our work addresses the open research problem of studying how to estimate privacy leakage from tabular DP-SDGs tightly.
We are the first to develop white-box MIAs against PrivBayes and MST, to instantiate and adapt (active) white-box MIAs~\cite{hayes2017logan,nasr2021adversary} to DPWGAN, and to consider specific worst-case neighboring datasets. 
}

\section{Discussion \& Conclusion}

\subsection{Summary}
This paper focused on tightly auditing (six) differentially private synthetic data generation (DP-SDG) implementations.
We analyzed the key factors affecting tightness, running several MIAs in different threat models and experimenting with worst-case datasets.
Our analysis shows that the privacy leakage of DP-SDGs can indeed be tightly estimated empirically, but only for strong adversaries and worst-case neighboring datasets.
In the process, we proposed novel white-box MIAs against PrivBayes and MST and presented an adaptation of Nasr et al.~\cite{nasr2023tight}'s gradient canary attack to DPWGAN.

Furthermore, our automated auditing procedure discovered DP violations in most DP-SDG implementations, including a new DP one in the DPWGAN implementation submitted to the NIST DP Synthetic Data Challenge. %
Overall, we are confident that our work will encourage more research into automated auditing tools so that DP-SDG implementations can be verified easily and at scale.

\subsection{The Importance of Automated Auditing}
\label{sec:auditing_importance}
Designing automated auditing tools is an important area of research as these enable researchers and practitioners to find bugs and violations of formal guarantees in real-world implementations.
Arguably, this is particularly relevant in the context of Differential Privacy (DP), as DP is increasingly used in the wild to protect the privacy of real-world users~\cite{dpdeployments,mcmahan2022federated,erlingsson2014rappor,appledp}, as well as citizens in critical settings like the U.S.~Census~\cite{abowd2022tda}.
This extends to differentially private synthetic data generation (DP-SDG) tools, which are being deployed to protect the data of sensitive populations like the individuals in Microsoft's human trafficking dataset~\cite{microsoftiom} or in healthcare settings~\cite{ucdavis}.
Bugs in these production systems break these protections and enable adversaries to learn sensitive information about end users~\cite{tang2017privacy,gadotti2022pool}.

This makes it crucial to audit algorithms and implementations as a systematic way to verify and guarantee the privacy of vulnerable groups in the wild.
To this end, our work showcases how manual ``inspection'' by experts to find DP violations %
might miss some subtle violations; overall, manual analysis may not be scalable, as each version of a released DP-compliant software will have to be verified individually.
Conversely, automated auditing can cover a wider range of violations and be included in continuous integration pipelines, thus reducing the potential for DP violations to be missed.
Indeed, our experiments show that our auditing procedure can \textit{automatically} find DP violations in DP-SDGs, including new ones that were previously missed.

\subsection{Powerful Threat Models}
\label{sec:powerful_threat_models}
As discussed, our auditing procedure goes beyond black-box threat models typically used in state-of-the-art MIAs against tabular synthetic data~\cite{stadler2022synthetic}, considering more powerful ones -- i.e., white-box, active white-box, and worst-case dataset attacks.
Naturally, the stronger the threat models, the stronger the assumptions in place.
In particular, white-box attacks are generally less practical to mount, as it is not always clear how the adversary can gain access to the final fitted generative model.
Arguably, the active white-box and worst-case dataset attacks may be even less practical -- e.g., the former assumes that the adversary can actively, yet possibly stealthily, manipulate model training.
On the other hand, as argued in~\cite{hayes2017logan}, white-box attacks can be considered practical when models are released following a data breach or when they are compressed/deployed to smartphones.

Nevertheless, we emphasize that the purpose of auditing is to ensure that the \textit{provably correct} privacy guarantees of DP are not ``lost'' in practice -- e.g., due to implementation bugs -- regardless of the threat model.
Furthermore, DP violations can sometimes result in realistic privacy leaks as well.
For instance, our work shows that the metadata violation leads to a membership inference attack, in the black-box setting, with an AUC of $\geq 0.95$ for PrivBayes (DS), PrivBayes (Hazy), MST (Smartnoise), and DPWGAN (Synthcity) (see Figure~\ref{fig:qb_vs_dcr}).
Similarly, prior work~\cite{appledp} has also demonstrated that sensitive information (e.g., skin tone or political orientation) can leak from DP algorithms when empirical privacy guarantees do not match the intended theoretical ones.
Finally, DP is, by design, a robust mathematical framework that provides privacy protections even against {\em worst-case} threat models, including the ones considered in this paper.

\subsection{Computational Cost of Auditing}
A potential concern with automated auditing is the computational cost incurred.
Not only are state-of-the-art auditing tools affected by the number of models that have to be built, but they also depend on the computational efficiency of the individual implementations.
For instance, in our experiments, it took 6.72s to generate a synthetic dataset from the (downsized) ADULT dataset with PrivBayes (Hazy) and more than 5x longer (39.0s) with PrivBayes (DS). %
Fitting the 10,000 models required for auditing took, respectively, 35 mins and 3 hours 24 mins for PrivBayes (Hazy) and PrivBayes (DS) by parallelizing the computation on a server with an Intel Xeon CPU with 32 2.20GHz cores and 128GB of RAM.
DPWGAN and MST took longer, with their NIST implementations taking 4 hours 8 mins and 14 hours 18 mins, respectively.

While we believe this is ultimately reasonable, reducing the number of models needed for auditing could be interesting for future work.
Incidentally, note that recent work~\cite{steinke2023privacy,galen2024oneshot} has presented one-shot auditing techniques (i.e., only using one model); however, these methods are specific to auditing differentially private stochastic gradient descent and do not provide tight empirical guarantees. %

\subsection{Limitations \& Future Work}
\label{sec:limitations}
Although our work succeeds in providing (almost) tight empirical estimates of privacy for the DP-SDG implementations studied, it is, naturally, not without limitations.

First, our auditing procedure requires thousands of synthetic datasets and models to be trained; this is due both to the use of shadow models, which trains a classifier on potentially thousands of samples, and the Clopper-Pearson confidence intervals limiting the maximum auditable $\varepsilon$.
For a given number of test observations, even when an adversary can perfectly distinguish between $\mathcal{M}(D)$ and $\mathcal{M}(D')$, i.e., $\alpha = \beta = 0$, the 95\% upper bounds $\overline{\alpha}$ and $\overline{\beta}$ are lower bounded.
As these bounds are used to calculate $\empeps$, this results in an upper bound on the $\empeps$ as well.

Second, we need millions of observations to audit at relatively large values of $\varepsilon$, such as $\varepsilon = 10$, using \approxdp~\cite{nasr2021adversary}.
While using credible intervals from~\cite{zanella2023bayesian} could improve on this, we find that, for 2,000 test observations, the difference in maximum auditable $\varepsilon$ is only 0.51.
Auditing with \gdp requires much fewer observations ($\approx 22$) but can only be applied to mechanisms that satisfy \gdp, thus excluding pure DP mechanisms like PrivBayes.

Recent work~\cite{steinke2023privacy,galen2024oneshot} propose techniques to audit DP discriminative models using only one trained model (``one-shot'').
However, they do not provide tight empirical guarantees, and it is not clear how they can be applied to generative models.
Therefore, %
we leave exploring these directions %
to future work.

In the future, we also plan to explore one or few-shot empirical privacy estimation of DP-SDGs and explore the deployment of our procedure into continuous integration pipelines.

\subsection{Ethics \& Disclosure}\label{sec:ethics}
Our work does not involve attacking live systems or private datasets.
In the spirit of responsible disclosure, in February 2024, we reported the five DP violations discussed in this paper to the respective library authors.
We offered to clarify, assist in fixes, and provide initial suggestions and recommendations.
We also refrained from making our findings public for at least 90 days from disclosure.

As of May 2024, only the authors of PrivBayes (Hazy) and DPWGAN (Synthcity) have responded to our disclosure.
The PrivBayes (Hazy) library now displays a privacy warning to users when it automatically learns the metadata from the dataset. 
Unfortunately, the latest version of DPWGAN (Synthcity) (v0.2.10) still contains both the metadata and PRNG reuse violations.
PrivBayes (DS), MST (Smartnoise), and DPWGAN (NIST) have not made any commits to their GitHub repository since then; thus, the violations are still present in the publicly available libraries.

\descr{Acknowledgments.} This work has been supported by the National Science Scholarship (PhD) from the Agency for Science Technology and Research, Singapore.

{\small
\bibliographystyle{plain}

}

\appendix
\section{Comparing Feature Sets for White-Box\\Attacks}
\label{app:feat_types}

\begin{figure}[t]
  \centering
  \includegraphics[width=\mywidth\linewidth]{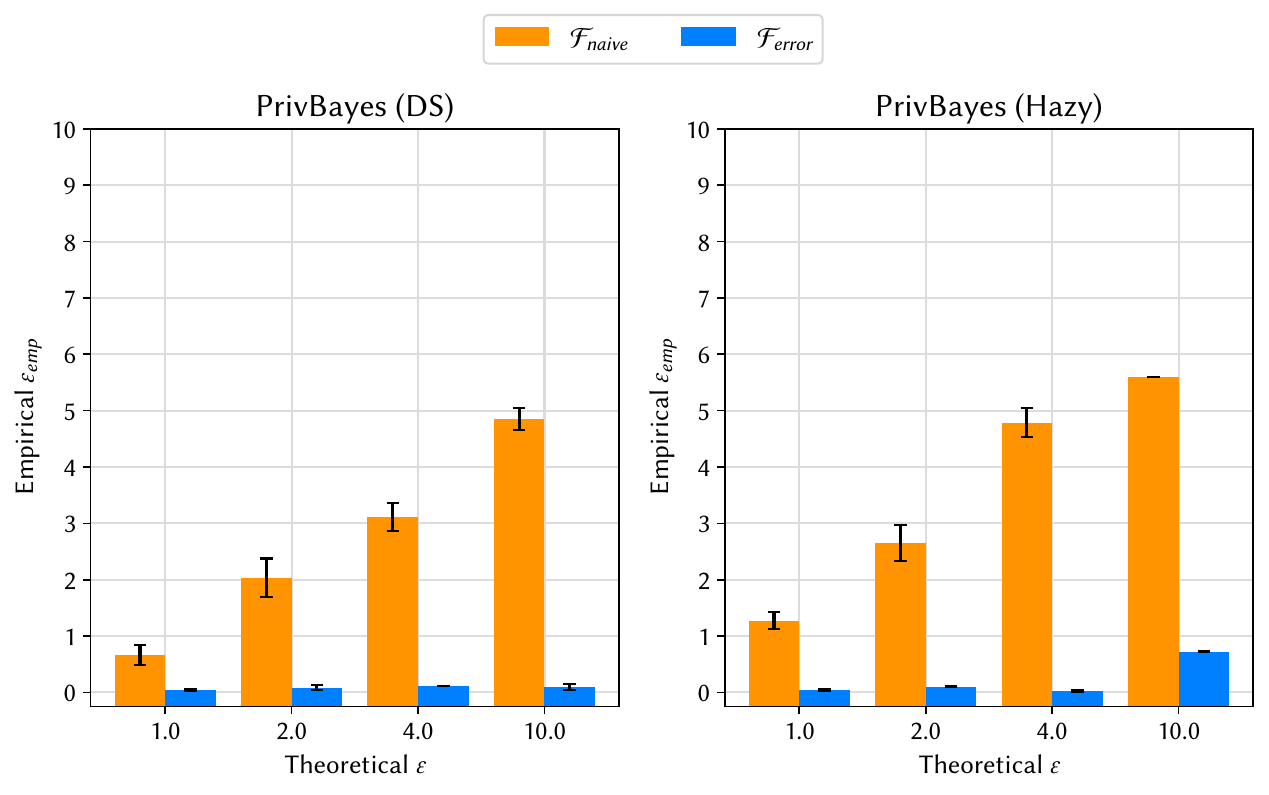}
  \caption{White-box auditing of PrivBayes implementations for different feature sets.} %
  \label{fig:feat_type_privbayes}
\end{figure}

\descr{PrivBayes.} In Figure~\ref{fig:feat_type_privbayes}, we plot the empirical $\empeps$ guarantees obtained when auditing PrivBayes using the implementation-specific worst-case dataset (see Appendix~\ref{app:compare_worstcase_wb}) but for different white-box features ($\mathcal{F}_{naive}$ and $\mathcal{F}_{error}$) that are extracted from the fitted generative models.
We find that the raw model parameters ($\mathcal{F}_{naive}$) result in much better guarantees than the error value feature set ($\mathcal{F}_{error}$) for all $\varepsilon$-s.

\begin{figure}[t]
  \centering
  \includegraphics[width=\mywidth\linewidth]{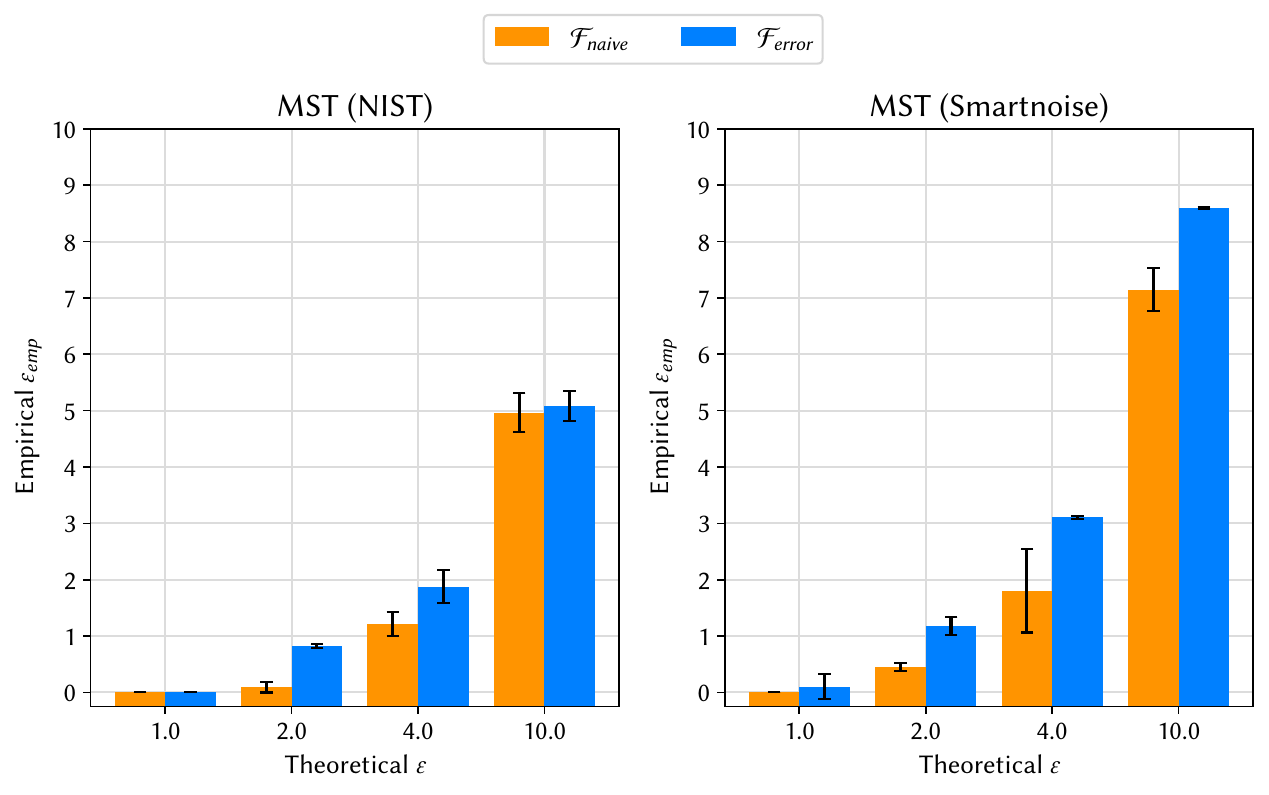}
  \caption{White-box auditing MST implementations for different feature sets.} %
  \label{fig:feat_type_mst}
\end{figure}

\descr{MST.} In Figure~\ref{fig:feat_type_mst}, we plot the empirical $\empeps$ guarantees obtained when auditing MST using the implementation-specific worst-case dataset (see Appendix~\ref{app:compare_worstcase_wb}) but for different white-box features ($\mathcal{F}_{naive}$ and $\mathcal{F}_{error}$) that are extracted from the fitted generative models.
We find that the error value feature set ($\mathcal{F}_{error}$) results in marginally tighter guarantees for all $\varepsilon$-s.

\section{Comparing Worst-Case Datasets for White-box Auditing}
\label{app:compare_worstcase_wb}

\begin{figure}[t]
  \centering
  \includegraphics[width=\mywidth\linewidth]{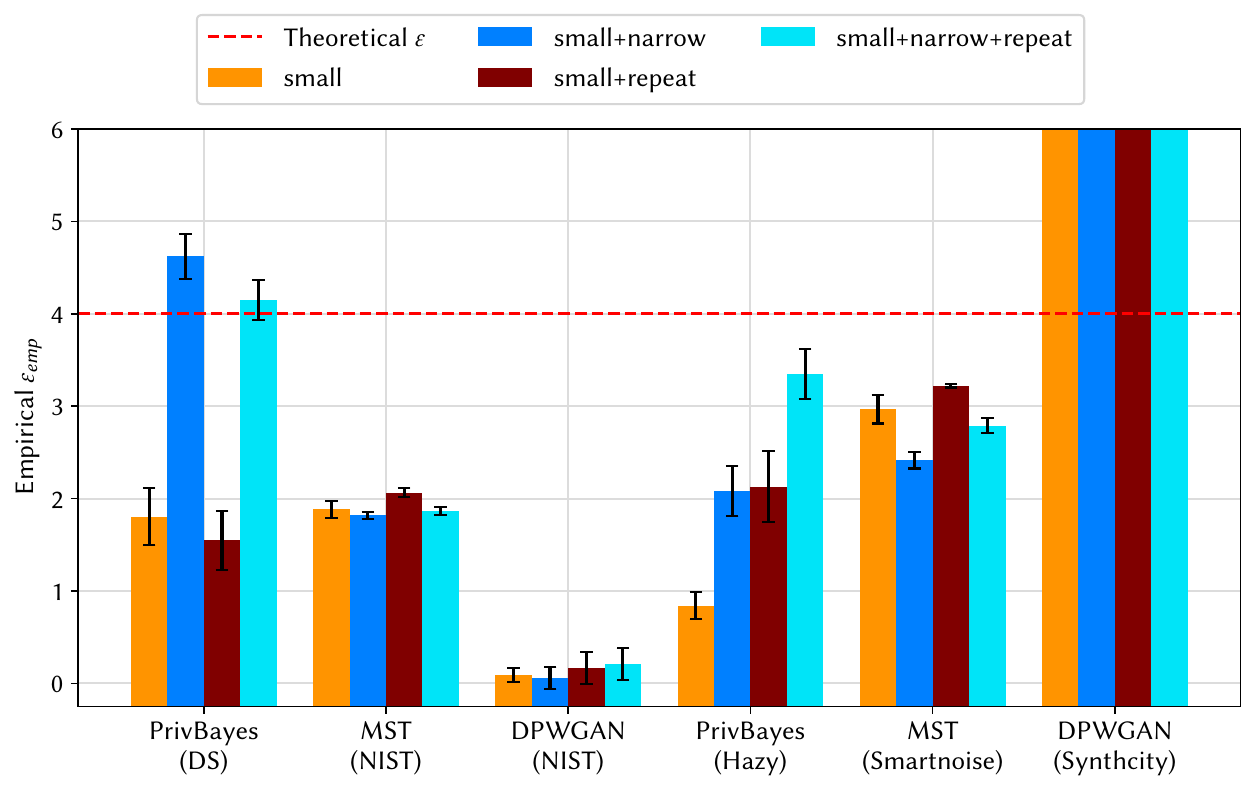}
  \caption{White-box auditing DP-SDG implementations at $\varepsilon = 4.0$ for different worst-case datasets.}
  \label{fig:compare_worstcase_wb}
\end{figure}

Figure~\ref{fig:compare_worstcase_wb} compares the empirical $\empeps$ guarantees obtained by the white-box attacks (specific to the DP-SDG implementation) for different worst-case datasets.
Similar to the black-box setting, we find that the worst-case dataset is implementation-dependent even for white-box attacks.

Specifically, for the PrivBayes (DS) and PrivBayes (Hazy) implementations, the \texttt{small+narrow} dataset and \texttt{small+narrow+repeat} dataset produce the tightest guarantees similar to the black-box setting.
On the other hand, for the MST (NIST), MST (Smartnoise), and DPWGAN (NIST), the \texttt{small+repeat} dataset produces the tightest guarantees.

\end{document}